\documentstyle[aps,preprint]{revtex}

\begin{document}
\author{O. Chmaissem$^{a},$ Y. Eckstein$^{b},$ and C.G. Kuper$^{b}$}
\address{$^{a}$Material Sciences Division, Argonne National Laboratory,\\
Argonne, IL 60439, U.S.A, \ and\\
Department of Physics, Northern Illinois University, DeKalb, IL 60115, U.S.A.%
\\
$^{b}$Department of Physics and Crown Center for Superconductivity, \\
Technion --- Israel Institute of Technology, \ 32000 Haifa, Israel}
\title{The Structure and a Bond-Valence-Sum Study of the 1-2-3 Superconductors (Ca$%
_{x}$La$_{1-x}$)(Ba$_{1.75-x}$La$_{0.25+x}$)Cu$_{3}$O$_{y}$ and YBa$_{2}$Cu$%
_{3}$O$_{y}$\\
}
\date{14.9.00}
\maketitle

\begin{abstract}
We present a study of the crystal structure of the 1-2-3 superconductor (Ca$%
_{x}$La$_{1-x}$)(Ba$_{1.75-x}$La$_{0.25+x}$)Cu$_{3}$O$_{y}$ (``CLBLCO''). \
Because of the presence of both La and Ba ions in the ``Ba'' layer, the
local symmetry of \ YBa$_{2}$Cu$_{3}$O$_{y}$ (``YBCO'') is lost. \ One can
no longer assume that an ``apical'' oxygen atom always lies strictly on the
line joining CuI and CuII ions, and a Pauling bond-valence calculation is
very useful as an adjunct to the Rietveld refinement of the positions of the
La and Ba ions. When a Rietveld analysis is performed allowing for the
possibility that the apical oxygen atoms may be slightly displaced along the 
$a$ or $b$ direction, the bond valence sums for both La and Ba are close to
the natural oxidation states of these ions.

We have also used the bond-valence-sum method to study the mean oxidation
states of the CuI and CuII ions in both CLBLCO and YBCO. \ Our results for
YBCO differ from some previously published results. \ Our main conclusion is
that, while the method may be useful in finding the charge concentration on
the CuII plane, it definitely does not yield the concentration of {\it mobile%
} holes, nor the concentration of superconducting charge carriers.


\end{abstract}

\section{{}{}Introduction}

Soon after the discovery\cite{chu} of YBa$_{2}$ Cu $_{3}$O$_{y}$\ (``YBCO'')
--- the first of the ``1-2-3'' high-temperature superconductors --- its
structure (Fig. 1) was reported\cite{structure}. \ Various substituents have
also been studied, but under the assumption that no substantial changes in
the structure occur. \ Besides illustrating the \ well-known 1-2-3
structure, Fig. 1 also defines our labelling convention for the distinct
copper and oxygen sites.

The critical temperature $T_{c}$ of cuprate superconductors can be varied by
changing the ``doping'' level ({\it i.e. }by varying the oxygen content \ $y$%
\ or by replacing a cation with another of dif{\hskip0pt}ferent valency). \
\ There is an ``optimum'' doping level, for which $T_{c}$ \ is a maximum. \
However, in the 1-2-3 series, overdoping by {\it only} varying the oxygen
content is usually limited to a very small decline in $T_{c}$ . \ 

This paper reports a study of an intriguing family\cite{knizhnik} of 1-2-3
materials (Ca$_{x}$La$_{(1-x)}$)(Ba$_{(c-x)}$La$_{(2-c+x)}$)Cu$_{3}$O$_{y}$
\ \ (``CLBLCO''). \ An attractive feature of CLBLCO is that, {\it regardless 
}of the values of $x$ and $y,$ it\ always crystallizes in the tetragonal
space group $P4/mmm$ $.$\ \ As indicated by the way in which we have written
the chemical formula, the lanthanum ions occupy two independent
crystallographic sites: the ``yttrium'' ($\frac{1}{2},\frac{1}{2},\frac{1}{2}%
)$ and the ``barium'' ($\frac{1}{2},\frac{1}{2},z)$ sites of YBCO. \ (But
note that in CLBLCO, the $z$ of La differs slightly from that of Ba .) \
These two La ionic positions will be denoted La$_{({\rm Y)}}$ and La$_{({\rm %
Ba)}}$ respectively. \ \ The tetragonality of the material implies that
there are no extended ``chains'' (such as exist in YBCO); the oxygen atoms
in the CuI layer are distributed randomly with respect to the $a$ and $b$
directions, and therefore the notation O$_{b}$ is inappropriate. \ We will
call these oxygen atoms O$_{\alpha }$, and similarly we will not distinguish
between O$_{pa\text{ \ }}$and O$_{pb}$ sites; we label them O$_{p}$. \ We
note that, for a given set of parameters, our recipe for preparing samples
always yields the same $T_{c};$ this implies that the different types of
unit cells ({\it e.g. }cells containing Ca and Ba, La and Ba, {\it etc.)}
are randomly distributed. This fact, together with the tetragonality of the
crystal, gives us confidence that the distribution of O$_{\alpha }$ atoms is
uniform (although there is no direct evidence).\ \ \ \ Because the ionic
radii of Ba ($R_{({\rm Ba}^{+2})}=1.34{\rm \AA })$ and La $(R_{({\rm La}%
^{+3})}=1.016{\rm \AA })$ are so very different, we might expect that there
will be some distortion of the lattice, and in particular\ that the O$_{c%
\text{ \ }}$and O$_{\alpha }$ atoms will suffer some lateral displacement. \
A study of this distortion will be one of the main topics of the present
paper.

The second attractive feature of CLBLCO is that it is possible to cover the
entire range from complete underdoping to complete overdoping by varying only%
{\it \ }the oxygen content $y$\ (in contrast to YBCO, which can be
significantly overdoped only by introducing some foreign cations).

The source for our analysis will be the neutron-diffraction data of
Chmaissem {\it \ et al.}\cite{chmaissem} \ The method will be a
``symbiotic'' one, using the techniques of Rietveld refinement\cite[ref3]
{rietveld} and Pauling bond-valence summation\cite{pauling}\cite{zach}\cite
{marez} (BVS). \ \ We shall also use the BVS technique to determine
separately the average oxidation states (``valence'') of the CuI and CuII
ions (since from the stoichiometry it is only possible to compute the global
average). \ We have made BVS calculations for YBCO, and disagree with
previously published results which purported to find the concentration of
mobile holes on the CuO$_{2}$ planes. \ We shall demonstrate that while this
procedure probably gives a reasonable approximation to the correct oxidation
states of the CuI and CuII layers, it does {\it \ not }give the
concentration of mobile charge carriers.

\section{Pauling's Bond Valence Sum Technique}

Pauling\cite{pauling}\cite{jansen} \ introduced the concept of fractional
valences, which he ascribed to the interionic bonds in an ionic crystal,
such that the sum of the bond valences (BVS) on an ion shall be equal to its
oxidation state. The bond valence between a cation and a neighboring anion
is usually well represented by exp\{$(r_{0}-r)/B$\}, where $B$ is a
universal constant having the value 0.37 {\AA , where } $r_{0}$ has a
characteristic value for a given cation--anion pair and where $r$ is the
interionic distance. Where an ion can exist with several dif{\hskip0pt}%
ferent oxidation states, the value of $r_{0}$ will depend on the oxidation
state. \ Table I shows all the relevant $r_{0}$ values (taken from Brown and
Altermatt\cite{brown-alt}).\ The idea is that the sum of these empirical
b{}ond valences about a given ion should agree with that ion's oxidation
state. Any significant discrepancy (say over 30\%) between the BVS and the
true oxidation state represents strain in the crystal, and may even indicate
that the assumed structure is incorrect. \ The BVS technique has sometimes
proved a useful tool\cite{zach} in checking crystal structures.

In \S 3, we first report the positions of the Ba and La$_{({\rm Ba)}}$ ions,
as found by a preliminary Rietveld refinement of the neutron-diffraction
data of Chmaissem {\it et al.}\cite{chmaissem}. \ \ We find two indications
that the resulting picture cannot be exact:

(a) When we\ calculate BVS's for these ions, we find BVS's for La$_{{\rm (Ba)%
}}$ $\ $below 2; these are manifestly unacceptable, since the oxidation
state of lanthanum is known unambiguously to be La$^{+3}$. \ This
discrepancy is an indication that this layer is indeed significantly
distorted.

(b) The presence of anomalously large atomic displacement parameters
(``adp's'') \ $U_{{\rm CuI}},$ $\ U_{{\sf O}_{\alpha }},$ and $U_{{\rm O}%
_{c}}$, for the CuI, O$_{\alpha }$, \ and O$_{c}$ ions\ respectively, tend
to confirm the above indication.

In \S 4, guided by the anomalous BVS values, we repeat the Rietveld
analysis, without the constraint that the O$_{\alpha }$\ and O$_{c}$ \ atoms
occupy their ``ideal'' positions (we contunue to rely on the fact that the
La and Ba ions in the ``Ba'' layer are distributed randomly). \ The new
Rietveld-refined structure is then tested by recalculating the BVS's for the
La \ and Ba ions. \ We show that the new structure is fairly satisfactory.

\ We have previously reported\cite{lt22} BVS calculations for the CuI and
CuII ions and the O$_{\alpha }$ and O$_{c}$ ions, both\ in YBCO and in
CLBLCO. \ The motivation for these calculations was to determine the
separate average oxidation states of the two types of Cu sites. \ However,
the global average of the BVS's of the CuI and CuII layers does not agree
with the global average oxidation state as determined by the stoichiometry.
\ \ This discrepancy arises because the interatomic distances (and therefore
also the BVS's) are constrained by the lattice structure; these internal
strains must somehow be taken into account. \ 

We will describe the methods proposed by Brown\cite{brown} and Tallon\cite
{tallon} to correct for the discrepancy. In particular, Brown calculates the
average oxidation state $2+p$\ of the CuII ions, \ and asserts that this $p$
is the concentration of Cu$^{+3}$ ions in the CuO$_{2}$ plane (after making
a correction for the strain), {\it i.e.} $p$ is the number of holes in this
plane. \ In Tallon's approach, the concentration of Cu$^{+3}$ ions is not
assumed to be the number of holes in the CuO$_{2}$ plane; \ he includes a
contribution from the BVS's of the oxygens to estimate the number $p$ of
holes in the plane. \ The two approaches do not agree; we describe both of
them, and discuss our preference. \ We also challenge the widely-held
conjecture that $p$ represents the concentration of {\it mobile }holes.

\section{\protect\bigskip Na{\"{i}}ve Bond Valence Sums for Calcium,
Lanthanum, and Barium}

\bigskip Table 2 is a specimen table\cite{table}, for calcium concentration $%
x=0.4$ and oxygen concentration $y=6.898$ giving the positions and adp's of
all the ions, as calculated by rietveld refinementfrom the
neutron-diffraction data of Chmaissem {\it et al.}\cite{chmaissem}, assuming
that the structure is undistorted. \ The lattice constants and the distances
are in Angstrom units, in all the tables, and the adp's \ in Tables 2 and 5
are in units of \AA $^{2}$. \ Table 3 \ is the full table of the relevant
interionic distances, calculated from the ionic coordinates. \ \ Using these
distances as a first approximation, the BVS's of Ca, La, Ba, and O were
calculated, and are shown in Table 4, as BVS$_{\text{Ca}}^{{\rm 0}},$ etc.
The results clearly show that distortion is present.

An example of the calculation is given, for $x=0.4$ and $y=6.898$:

\begin{eqnarray}
{\rm BVS}_{{\rm La}_{({\rm Ba})})}^{0} &=&\sum {\rm exp}((r_{0}-r)/B \\
&=&4\;{\rm \exp [\{}r_{0}({\rm La)-}r({\rm La}_{{\rm (Ba)}}~{\rm to}~{\rm O}%
_{p})\}{\rm /}B]  \nonumber \\
&&+4\;{\rm \exp [\{}r_{0}({\rm La)-}r({\rm La}_{{\rm (Ba)}}~{\rm to}~{\rm O}%
_{c}){\rm \}/}B]  \nonumber \\
&&+2(y-6){\rm \exp [\{}r{\rm _{0}(La)-}r({\rm La}_{{\rm (Ba)}}~{\rm to}~{\rm %
O}_{\alpha })\}{\rm /}B]  \nonumber \\
&=&4\;{\rm exp\{(}2.172-2.721)/0.37\}+4\;{\rm exp\{(}2.172-2.797)/0.37\} 
\nonumber \\
&&+0.852\;{\rm exp}\{(2.172-3.055)/0.37\}  \nonumber \\
&=&1.811
\end{eqnarray}

$\smallskip \ $

\medskip We see that the BVS of La$_{({\rm Ba)}}$ is quite unacceptable,
although the BVS's of La$_{({\rm Y)}}$ and Ca are reasonable. \ We attribute
this anomaly for La$_{({\rm Ba)}}$ to the distortion induced by the great
discrepancy between the La and Ba ionic radii. A further indication that the
lattice is distorted is the fact that the adp's found by Chmaissem {\it et
al.}\cite{chmaissem}{\it \ }for some of the oxygen atoms{\bf \ }are
anomalously large. \ They can be improved significantly by assuming that
they are not isotropic, but have elliptical symmetry. \ However, we do not
believe that these adp's are thermal in origin, but rather reflect the
possibility that the positions of the O$_{\alpha }$ and O$_{c}$ ions depend
on the nature of their neighbors, and may suffer static displacements.\ 

\section{Improved Bond Valence Sums for Calcium, Lanthanum, and Barium}

We should expect the displacements of O$_{c}$ and O$_{\alpha }$ ions to
depend on their environment. \ Each O$_{c}$ has four closest neighbors,
which may be La$^{+3}$ or Ba$^{+2}$. These ions have very different radii ($%
R_{({\rm La}^{+3})}=1.016$ \AA ; $R_{({\rm Ba}^{+2})}=1.34$ \AA ). \ We
shall assume that the La and Ba are distributed randomly. There are six
possible environments for an O$_{c}$, as illustrated in Fig. 2. \ In cases
A, B, and C, the symmetry implies that the O$_{c\text{ }}$oxygen atom will
not suffer any sideways displacement (but it can be displaced longitudinally
to achieve favorable bond lengths). \ However, because the ionic radius of
La is so much smaller then that of Ba, the O$_{c}$ in cases D, E, and F will
clearly be displaced away from Ba ions and towards La ions.{\bf \ }\ We also
note that cases A and B are nondegenerate, that case C is doubly degenerate,
and that cases D, E, and F are each fourfold degenerate. \ \ \ Let the
number of Ba ions per formula unit be 2$b$, and the number of La$_{({\rm Ba}%
\text{) }}\ $ ions be 2$l,$ (where, from the chemical formula, $l+b=1).$ 
{\bf \ }Because the distribution is assumed to be random, the percentage of
displaced O$_{c}$ ions in each of the six configurations can be calculated.
\ Thus the total probability $q_{1}$ of the O$_{c}$ not being displaced is

\[
q_{1}=l^{4}+b^{4}+2b^{2}l^{2}, 
\]

\noindent the probability \ $q_{2}$ that O$_{c}$ is displaced \ diagonally \
is

\[
q_{2}=4(l^{3}b+lb^{3}), 
\]

\noindent and the probability $q_{3}$ of displacement along the $a$ or $b$
crystallographic directions is

\[
q_{3}=4b^{2}l^{2}. 
\]

\noindent Note that $q_{1}+q_{2}+q_{3}=1,${\it \ i.e.} we have taken all
possible configurations into account.

\ The Rietveld refinement was repeated, allowing the \ O$_{c}$ ions to
occupy slightly different positions $r({\rm O}_{c(1)}),r({\rm O}_{c(2)}),r(%
{\rm O}_{c(3)}),$ where $r({\rm O}_{c(1)})$ lies along the $c$-axis, but $r(%
{\rm O}_{c(2)})$ is displaced diagonally from the $\ c$-axis, and $r({\rm O}%
_{c(3)})$ is displaced along an $a$ or $b$ direction from the $\ c$-axis
(Table 3). \ The relative occupancy of the three sites are given by the
configuration-dependent probabilities \ $q_{1},$ $q_{2}$ and $q_{3}.$

Analogously, the O$_{\alpha }$ ions can also suffer displacements depending
on their environments. \ The calculation here is somewhat simpler, since
there appears to be a tendency for the La$_{({\rm Ba)}}$ ions to dimerize,
i.e. if there is an La ion at position ($\frac{1}{2},\frac{1}{2},\zeta ),$
it is probable that there is another one at ($\frac{1}{2},\frac{1}{2},-\zeta
).$ \ This tendency is indirectly confirmed by the Rietveld neutron
refinements. \ The presence of a lanthanum dimer should leave the O$_{\alpha 
\text{ \ }}$ion in the basal plane. Indeed the $z$ coordinates of the O$%
_{\alpha }$'s refined to zero, confirming that there is no displacement
along the $z$ direction and supporting our view that the La ions indeed form
dimers. \ However, when the O$_{\alpha }$'s are allowed to occupy the two
sites ($\frac{1}{2},0,0)$ and ($\frac{1}{2},\eta ,0)$ --- but are
constrained\ to be confined to the basal plane --- the adp is greatly
reduced. \ Allowing only ($\frac{1}{2},0,0)$ sites yields adp's in the range
from \ $U_{{\rm O}_{_{\alpha }}}=$ \ 4 to 7, but allowing the additional
site ($\frac{1}{2},\eta ,0)$ reduces them to the reasonable range from $U_{%
{\rm O}_{_{\alpha }}}=$ \ 1.5 to 2. Some correlation was seen between the
position, the site occupancy, and the adp's for the two sites, and a final
refinement was made keeping this adp fixed at its average value, $U_{{\rm O}%
_{_{\alpha }}}=1.8$. \ This refinement is the source\cite{table} for new
positions and occupancies --- see the specimen Table 5, which replaces Table
2.\ Table 6 is the full table of the interatomic distances, calculated from
the revised coordinates; these new distancesare used in the subsequent
calculations.

\ The BVS calculations for Ca and La$_{({\rm Y)}}$ are straightforward, and
proceed exactly as in \S 3. \ However, the calculations for Ba and La$_{(%
{\rm Ba)}}$, require some care.\ \ The main new feature in calculating the
Ba and La$_{({\rm Ba)}}$ sums is that the O$_{c}$ environment of a given
cation will depend on the three other cation neighbors of that O$_{c}$. \
Thus, if we focus our attention on a given La ion, with its four O$_{c}$
neighbors, the probability $Q(1,{\rm La}_{{\rm (Ba)}})$ that one of its O$%
_{c}$ neighbors is undisplaced is (from cases B and C, Fig. 2)

$Q(1,{\rm La}_{{\rm (Ba)}})=l^{3}+b^{2}l,$

\noindent where $b$ and $l$ are the concentrations of Ba and La respectively.

\bigskip Cases C and D yield a total probability {\rm of} $\ {\rm 3}%
bl^{2}+b^{3},$ \noindent which we split into the two parts

$Q(2,{\rm La}_{{\rm (Ba)}}{\rm )=}bl^{2}+b^{3},$

\noindent corresponding to the case where the O$_{c}$ is ``pushed towards''
the given La, and

$Q(3,{\rm La}_{{\rm (Ba)}}{\rm )=2}bl^{2},$

\noindent where the displacement of O$_{c}$ is orthogonal to the undistorted
La---O$_{c}$ bond direction. Finally, the probability $Q(4,{\rm La}_{{\rm %
(Ba)}}{\rm )}$of displacement along an $a$ or $b$ direction is:

$Q(4,{\rm La}_{{\rm (Ba)}}{\rm )=}2bl^{2}.$

\noindent The $Q^{\prime }$s for Ba are obtained similarly, by interchanging 
$l$ and $b$.

As an example, we show the recalculated BVS for La$_{{\rm (Ba)}}$ for $x=0.4$
and $y=6.926$, (which supersedes Eq. (2)): \ 
\begin{eqnarray}
{\rm BVS}_{{\rm La}_{{\rm (Ba)}}} &=&\sum {\rm exp}\{(r_{0}-r)/B\}  \nonumber
\\
&=&{\rm 4\;\exp [\{}r_{0}({\rm La})-r({\rm La}_{{\rm (Ba)}}{\rm ~to~O}%
_{p})\}/B]  \nonumber \\
&&+4\sum_{s=1}^{4}{\cal Q}_{s}(s,{\rm La}){\rm \exp [\{}r_{0}({\rm La})-r(%
{\rm La}_{{\rm (Ba)}}{\rm ~to~O}_{c(s)})\}/B]  \nonumber \\
&&+2\sum_{t=1}^{2}n_{t}{\rm \exp [\{}r_{0}({\rm La})-r({\rm La_{(Ba)}~to~O}%
_{\alpha (t)})\}/B]  \nonumber \\
&=&4\;[{\rm exp\{(}2.172-2.937)/0.37\}\ \ \ \   \nonumber \\
&&+0.182\;{\rm exp\{(}2.172-2.754)/0.37\}+0.379\;{\rm exp\{(}%
2.172-2.380)/0.37\}  \nonumber \\
&&+0.143\;{\rm exp\{(}2.172-2.779)/0.37\}+0.296\;{\rm exp\{(}%
2.172-2.457)/0.37\}]  \nonumber \\
&&+2[0.516\;{\rm exp}\{(2.172-2.831)/0.37\}+0.408\;{\rm exp\{}%
(2.172-2.378)/0.37\}]  \nonumber \\
&=&2.821
\end{eqnarray}

\ $\ $

\medskip This example illustrates the improvement which the revised Rietveld
refinement has made in the valence of La.

\section{Average Valence of Copper in the Copper-I and Copper-II layers}

All the non-copper ions have well-defined oxidation states; only that of the
copper is variable, taking values ranging from +1 to +3 depending on the
oxygen content $y$. (Although, by its definition, the oxidation state of an
atom can take only integer values, the {\it average} oxidation state of Cu
on the CuI or CuII layer will, in general, not be an integer.)

The dif{\hskip0pt}ficulty in assigning oxidation states to the Cu ions in
1-2-3 materials is that the stoichiometry can only give a {\it global }%
average oxidation state. Pauling's empirical BVS method was first applied by
de Leeuw {\it et al.}\cite{leeuw}, and Cava {\it et al.}\cite{cava}, and it
has become popular\cite{tallon} to rely on this method, in order to assign
separate average oxidation states to the ``chain'' and ``plane''copper
layers (CuI and CuII layers respectively). \ 

\smallskip The motivation for using BVS's is the belief that they yield good
approximations to the actual oxidation states. \ Following Brown\cite{brown}%
, we calculate BVS's for copper in both the CuI and CuII layers; our
expectation is that the appropriately weighted average of the CuI and CuII
BVS's should yield the true average oxidation state. \ Let us assume that a
fraction $\xi $ of the Cu ions in one of the layers are Cu$^{+3}$, so that $%
(1-\xi )$ are Cu$^{+2}.$ The average charge (oxidation state) per Cu ion in
this layer is thus

\[
V_{{\rm Avg}}=3\xi +2(1-\xi )=2+\xi . 
\]

\noindent This should be equal to $\xi V^{(+3)}+(1-\xi )V^{(+2)},$ where $%
V^{(+3)}=({\rm BVS}_{{\rm Cu}^{+3}})$, {\it i.e }the BVS, calculated using $%
r_{0({\rm Cu}^{+3}{\rm )}}=1.73$ \AA , and $V^{(+2)}={\rm (BVS}_{{\rm Cu}%
^{+2}})$. \ Hence, solving for $\xi $, we have ({\it cf.} Brown\cite{brown}):

\begin{equation}
\xi =(V^{(+2)}-2)/(V^{(+2)}+1-V^{(+3)}).
\end{equation}

\ Our procedure is thus: (a) calculate $V^{(+2)}$ and \ $V^{(+3)}$, for the
CuI layer, and find $\xi _{{\rm I}}$ ({\it i.e.} the concentration of Cu$%
^{+3}$ ions in this layer); \ (b) similarly calculate $V^{(+2)}$ and \ $%
V^{(+3)}$ for the CuII layer, and find $\xi _{{\rm II}}$, \ (c) find the
global average (remembering that there are two CuII layers and only one CuI
layer in the unit cell): 
\begin{equation}
V_{{\rm Global}}^{{\rm Avg}}=(V_{{\rm CuI}}^{{\rm Avg}}+2V_{{\rm CuII}}^{%
{\rm Avg}})/3=(6+\xi _{{\rm I}}+2\xi _{{\rm II}})/3.
\end{equation}

As an example, we describe the detailed calculation of $V_{{\rm CuII}%
}^{(+2)} $ for CuII, with $x=0.4$ and $y=6.926:$

\begin{eqnarray}
V_{{\rm CuII}}^{(+2)} &=&4\;{\rm \exp [\{}r_{0}({\rm Cu}^{+2})-r({\rm %
CuII~to~O}_{p})\}/B]  \nonumber \\
&&+\sum_{u=1}^{3}q_{u}{\rm \exp [\{}r_{0}({\rm Cu}^{+2})-r({\rm CuII~to~O}%
_{c(u)})\}/B] \\
&=&4\;\exp \{{\rm (1.679-1.946)/0.37\}+0.315\exp \{(1.679-2,292)/0.37\}} 
\nonumber \\
&&+{\rm 0.492\exp \{(1.679-2.322)/0.37\}+0.193\exp \{(1.679-2.336)/0.37)\}} 
\nonumber \\
&=&2.123
\end{eqnarray}

$.$

The calculation of \ $V_{{\rm CuII}}^{(+3)}$ proceeds similarly, merely
replacing $r_{0}({\rm Cu}^{+2})=1.679$ \ by $r_{0}({\rm Cu}^{+3})=1.730,$
giving $V_{{\rm CuII}}^{(+3)}=2.437,$whence $\xi =0.180$, and the average
BVS of CuII is $V_{{\rm CuII}}^{{\rm Avg}}=2.180$. \ \ \ 

The calculations for CuI are similar, using the distances $\ r({\rm CuI~to~O}%
_{\alpha (1)})$ , \ $r({\rm CuI~to~O}_{\alpha (2)})$, $r({\rm CuI~to~O}%
_{c(1)})$, $r({\rm CuI~to~O}_{c(2)})$, and $r({\rm CuI~to~O}_{c(3)}),$ and
the appropriate weight factors $n({\rm O}_{_{\alpha }(u)})$ and $q_{u}$. \
The result is that $V_{{\rm CuI}}^{{\rm Avg}}=2.079$.

The global average BVS is \ 
\begin{equation}
V_{{\rm Global}}^{{\rm Avg}}=(V_{{\rm CuI}}^{{\rm Avg}}+2V_{{\rm CuII}}^{%
{\rm Avg}})/3=2.146.
\end{equation}

The full results are given in Table 8. \ Note that\ these averages do not
agree with the average oxidation states, as calculated from the
stoichiometry of the material. (For the example above, the global average
from stoichiometry is 2.201.) \ This discrepancy --- which we interpret as
an indication that bond lengths are constrained by the crystal structure ---
will be discussed in \S 8.

\section{Bond Valence Sums for the Oxygen Ions}

The calculations for oxygen are not completely straightforward. The
principal complication arises from the fact that the Cu ions may be in
either of the states Cu$^{+2}$ and Cu$^{+3}$, and therefore we must use the
mixing ratios $\xi _{{\rm I}}$ and $\xi _{{\rm II}}$ as weights in finding
the contributions from the Cu ions. \ \ Many previously published papers\cite
{tallon}\cite{cava}\cite{karpin}\cite{errors} did not make use of the $\xi
^{\prime }$s, but made the assumption, {\it de fa{\rm c}to,} that both $\xi
_{{\rm I}}$ and $\xi _{{\rm II}}$ were zero.

\smallskip We illustrate the calculation by the example of O$_{c(2)}$ \ for $%
x=0.4$ and $y=6.926$:

\begin{eqnarray}
({\rm BVS)}_{{\rm O}_{c(2)}} &=&\xi _{{\rm I}}{\rm \exp [-\{}r{\cal (}{\rm %
CuI~to~O}_{c(2)})-r_{0}({\rm Cu}^{3+)}\}/B]  \nonumber \\
&&+\xi _{{\rm II}}{\rm \exp [-\{}r{\cal (}{\rm CuII~to~O}_{c(2)})-r_{0}({\rm %
Cu}^{3+)}\}/B]  \nonumber \\
&&+(1-\xi _{{\rm I}}){\rm \exp [-\{}r{\cal (}{\rm CuI~to~O}_{c(2)})-r_{0}(%
{\rm Cu}^{2+)}\}/B]  \nonumber \\
&&+(1-\xi _{{\rm II}})\exp {\rm [-\{}r{\cal (}{\rm CuII~to~O}_{c(2)})-r_{0}(%
{\rm Cu}^{2+)}\}/B]  \nonumber \\
&&+\{b^{2}l/(b^{2}l+bl^{2})\}\;{\rm \exp [-\{}r({\rm Ba~to~O}_{c(2)})-r_{0}(%
{\rm Ba})\}/B]  \nonumber \\
&&+\{bl^{2}/(b^{2}l+bl^{2})\}\;{\rm \exp [-\{}r({\rm La}_{{\rm (Ba)}}{\rm %
~to~O}_{c(2)})-r_{0}({\rm La})\}/B]  \nonumber \\
&=&1.929
\end{eqnarray}

The full results for the oxygen BVS's are shown in Table 9.

\section{YBCO}

Cava\ \ {\it et\ al}.\cite{cava} have\ found\ the\ locations of\ the\ atoms
in YBCO \ by\ neutron\ diffraction,\ \ and\ have\ used\ their\ data\ to\
calculate\ the\ BVS's of all the ions present. \ In some respects the
situation is much simpler here than for CLBLCO. \ The O$_{c}$ and O$_{\alpha
}$ ions are never displaced, so that we do not have O$_{c(2)},$ O$_{c(3)}$
or O$_{\alpha (2)}$ present. \ However, there is a complication with respect
to the copper ions. \ When $y$ is close to 6, the average oxidation state in
the CuI layer (chain layer) can be less than 2, {\it i.e.} we may think of
the layer as containing a mixture of Cu$^{+}$ and Cu$^{+2}$. But for rather
larger $y$, the average oxidation state will be greater than 2, and the
layer will be a mixture of Cu$^{+3}$ and Cu$^{+2}$. (The assumption is made
that Cu$^{+}$\ and Cu$^{+3}$cannot coexist in any one Cu layer.) \ To allow
for the possibility that a mixture of Cu$^{+2}$ and Cu$^{+}$is present in a
layer, we introduce a new mixing ratio \ $\xi ^{(1)}$, which is easily shown
(by analogy to the calculation of $\xi ,$ see Eq. (4) ) to be: 
\begin{equation}
\xi ^{(1)}=(V^{(+2)}-2)/(V^{(+2)}-1-V^{(+1)}),
\end{equation}
where $V^{(+1)}$ is the BVS for Cu$^{+}$ ({\it i.e.} the BVS calculated with 
$r_{0}=1.60$). \ If Cu$^{+}$ is present, and not Cu$^{+3}$, we will find
that \ $\xi ^{(1)}$ is positive, and $\xi $ is negative, while when Cu$^{+3}$
is present, $\xi $ is positive and $\xi ^{(1)}$ negative. \ We therefore
always use whichever of the $\xi $'s is positive.\ This complication never
arises in CLBLCO, since there $y$ is never less than about 6.4, and no Cu$%
^{+}$ will be present; both the CuI and CuII layers will be mixtures only of
Cu$^{+2}$ and Cu$^{+3}$.

There are two reasons for repeating the BVS calculations for YBCO: \
Firstly, many of the published calculations\cite{tallon}\cite{cava}\cite
{karpin}\cite{errors} contain a conceptual error, which we wish to correct.
In calculating oxygen BVS's, these papers make the tacit assumption that
only Cu$^{+2}$ is present, {\it i.e. }that $\ \xi =\xi ^{(1)}=0$\ . \ And
secondly, we hope to show that the comparison between YBCO and CLBLCO can
help to clarify the question of the charge on the CuII plane. \ Moreover, we
discuss the question as to what fraction of that charge is mobile; and we
test whether there is a universal relation between the maximum value of $%
T_{c}$ and the concentration of holes for all high-$T_{c}$ materials (as
proposed by Tallon\cite{tallon}). \ \ We have recalculated the BVS's from
the data of Cava {\it et al.}, and were able to confirm their results for
copper, barium, and \ yttrium. \ However, we could reproduce their results
for oxygen only under the\ assumption\ that\ for\ this\ calculation\ $\xi
=\xi ^{(1)}=0,$\ ({\it i.e}. that all the copper ions in the CuO$_{2}$\
plane are Cu$^{+2}$).

\section{The Concentration of Charges in the Copper-I and Copper-II Layers}

\smallskip Historically, the BVS technique was introduced as an aid to
determining crystal structures\cite{zach}. \ It is clear from our analysis
that it is indeed valuable in clarifying details of the structure of CLBLCO.
\ However, it has also become popular recently to use the BVS method in an
attempt to find the distribution of electric charge in the high-$T_{c}$
cuprate materials. \ It is almost universally agreed that the
superconductivity of the cuprates resides on their CuO$_{2}$ planes. \ Can
BVS calculations really throw any light on the question of the charge
distribution on these planes?

\ \ As the prescription is empirical, it is not obvious how to interpret the
BVS results. \ Two attempted interpretations have been discussed. \ In one
approach, Tallon \cite{tallon} interprets the BVS of a given ion as giving
directly the {\it actual charge }on the ion{\it . }\ He calculates both the
average BVS of the CuII ions and the BVS of the O$_{p}$'s and defines:{\bf \ 
} 
\begin{equation}
V_{-}=(2+{\rm BVS}_{{\rm CuII}}-{\rm BVS}_{{\rm O}_{pa}}-{\rm BVS}_{{\rm O}%
_{pb}}),
\end{equation}
which, according to his interpretation, should be the concentration of holes
in the CuII plane. \ 

The second approach, by Brown\cite{brown}, uses only BVS$_{{\rm CuII}},$%
after making a correction for the internal strain. \ Brown considers that
the $\xi $'s are supposed to give directly the excess over 2 of the average
copper valence, and hence they should describe the concentration of holes in
the various Cu layers. \ The discrepancy between the global average BVS of
copper and the stoichiometric average copper valence is a consequence of the
strain in the lattice. \ It is not {\it a priori }\ obvious that the strain
correction to the Cu I and CuII layers are equal, but in the absence of any
better criterion we assume their equality. \ Thus the corrections which we
make to the CuI and CuII \ oxidation states are just{\it \ }$(V_{{\rm Global}%
}^{{\rm Stoich.}}-V_{{\rm Global}}^{{\rm Avg}}).$\ In the example of eqs.
(6,7), ($y=6.926,\;x=0.4),$we found (see Table 8b) that $V_{{\rm Global}}^{%
{\rm Avg}}=2.146$, while $V_{{\rm Global}}^{{\rm Stoich.}}=2.201.$\ \
Following Brown's prescription, we add the difference to each Cu layer,
giving the {\it corrected} average valences 
\begin{eqnarray}
{\cal V}_{C{\rm uI}}^{{\rm Avg}} &=&2.079+0.054=2.133,  \nonumber \\
{\cal V}_{C{\rm uII}}^{{\rm Avg}} &=&2.180+0.054=2.234,
\end{eqnarray}
and hence the net charge on the CuO$_{2}$ plane is 0.234. \ (Note that the
value in Table 9 --- namely 0.231 --- differs slightly from this number
because in the Table we made a linear least-square fit of the correction as
a function of $y.$

\smallskip \noindent \noindent \noindent Of course, if we knew how to
correct exactly for the strain, the two methods should agree (since, by
definition, the oxidation state of oxygen is exactly 2). \ Since the method
is largely empirical, it is impossible to say, {\it a priori, }which
approach is to be preferred. \ In the next section, we give reasons why, in
our opinion, Brown's approach is preferable

\section{Discussion}

One of the main reasons for the wide interest in BVS studies of the cuprates
was the hope \ that the results would be a help in understanding their
superconductivity. \ In particular, \ it is important for our understanding
of high-$T_{c}$ superconductivity to estimate the concentration of mobile
holes. \ \ In La$_{2-x}$Sr$_{x}$CuO$_{4},$ which does not have a ``chain''
layer, the average oxidation state of \ copper is determined by the
stoichiometry, to be $2+x$, since the oxidation states of La and Sr and
oxygen are unambiguously known (+3, +2 , and $-2$ respectively) . \ \ Hence $%
V_{-}=x,$ $(=0.16$ at optimal doping --- {\it i.e. }at maximal $T_{c})$, \
and this $x$ is taken to give directly the concentration of mobile holes. \
However, in the 1-2-3 materials, the stoichiometry can only give the {\it %
global }average oxidation state of the copper ions, and not the separate
valuers for the CuI and CuII. \ Moreover, it is generally believed that the
mobile charge carriers reside in the CuO$_{2}$ planes, and that in the 1-2-3
materials the chains serve primarily as a reservoir of charge; doping is
effected by transfer of charge from the CuI layer to the CuII layers.

The principal motivation for performing BVS calculations for copper was the
hope that they would enable one to find the average oxidation state of the
CuII layer, and hence, by analogy with La$_{2-x}$Sr$_{x}$CuO$_{4},$ the
concentration of mobile carriers. \ \ Tallon\cite{tallon}\cite{tallon-coo},
who followed Cava {\it et al.}\cite{cava} \ in taking \ $\xi =0$, found that
at optimum doping, $V_{-}^{{\rm YBCO}}\simeq 0.16${\bf .} \ This unfortunate
coincidence led to a widespread belief that this value 0.16 was universal
for optimally-doped high-$T_{c}$ cuprates. \ \ Furthermore, Tallon noted
that in the calculation of $V_{-},$ the CuII--O$_{p}$ bonds play no role, as
their contribution to BVS$_{{\rm CuII}}$ and to BVS$_{{\rm O}_{p}}$ exactly
compensate, and that therefore only the CuII--O$_{c}$ bonds are relevant. \
(This may have been a reason for the belief that the apical oxygens were
critically important.) \ Karpinnen and Yamauchi\cite{karpin}, relying on
this compensation but once again taking $\xi =0$, found $V_{-}=0.99.$ \ The
discrepancy arises from the inconsistent use of $\xi =0$, in calculating the
BVS of the oxygen and copper. \ When the correct value of $\xi $ is used
consistently, the two methods agree, and give $V_{-}=0.105$ for optimally
doped YBCO\cite{lt22}, \ forcing us to abandon belief in the existence of a
universal \ value.

We can make a further claim --- namely that although the BVS method can give
the total \ charge concentration on the CuII layer, it does {\it not} give
the mobile charge. \ The CLBLCO family (Ca$_{x}$La$_{(1-x)})($Ba$%
_{(1.75-x)}) $La$_{(0.25-x)})$CuO$_{y}$ has the useful property that the
average oxidation state of Cu is $\frac{1}{3}(2y-7.25),$ independent of the
Ca concentration $x$. \ For two different values of $x$, it was found\cite
{danny} that in the underdoped regime, samples with the same $T_{c}$ but
with different $y$ have the same resistivity $\rho $ and the same
thermoelectric power ${\cal S}.$ \ Although there is no adequate theory for
the thermoelectric power of the high-$T_{c}$ materials, we feel that we may
safely assume that the equality of $\rho $ and of ${\cal S}$ for equal $%
T_{c} $ means that they have the same mobile-carrier concentration, and also
from Uemura's\cite{uemura} $\mu $sr results, we know that samples of YBCO
with the same $T_{c}$ also have the same concentration of Cooper pairs. \
However, from the BVS calculations, our samples of CLBLCO with equal $T_{c}$
do {\it not }have the same {\it total} charge concentration. \ \ This surely
means that they have about the same concentration of mobile holes, although
they clearly do not have the same values of either $p_{{\rm Tallon}}$ or \ $%
p_{{\rm Brown}}$ (see Table 9). \ Not all the charges are mobile!

Another interesting new result for CLBLCO is that the average oxidation
states of CuII (calculated by Brown's method) for different $x$ but for the
same $y$ are equal (Fig.3). \ \ At present we have data for only two values $%
x=0.1$ and $x=0.4,$ but provided that this result is not accidental, it
means that CLBLCO is fully charge-compensated with respect to Ca and La; not
only is the {\it global} average oxidation state of copper independent of $x$%
, but the average oxidation states of the CuI and CuII layers are {\it %
separately} independent of $x$. \ This means (see Fig. 4) that in CLBLCO the
plot of $T_{c}$ versus $p_{{\rm Brown}}$ is essentially the same as its plot
against the oxygen concentration $y$. \ (Note that this will not be true for 
$p_{{\rm Tallon}}$; this is one of our reasons for tentatively preferring \ $%
p_{{\rm Brown}},)$ \ The fact that the two Ca concentrations ($x=0.1$ and $%
x=0,4)$ have such different $T_{c}$'s \ for the same value of $p$
strengthens our view that some of the holes are not mobile.

\smallskip The fraction of the hole concentration residing on the CuII layer
is $\ {\cal C=}2p/(V_{{\rm Global}}^{{\rm Avg}}-2).$ \ Zhu and Tafto\cite
{zhu}, using a novel electron-diffraction method, have been able to measure $%
{\cal C}$ directly for YBCO. \ They find that, close to optimal doping. \ $C$
$=$ $0.76\pm 0.08$, \ \ In Table 10, we give the values of $p_{{\rm Tallon}%
}^{\xi =0}$, $p_{{\rm Tallon}}^{\xi \neq 0},$ and $p_{{\rm Brown}}$ for
YBCO, \ (from ref. [16]) \ together with the values of \ ${\cal C}$\ ,
calculated from {\bf \ }these \ $p$'s. \ We see that $p_{{\rm Brown}}$ is in
fair agreement with the Zhu--Tafto results, while neither of the other
estimates of ${\cal C}$ is at all close. \ 

\ \ Note that the present values of $p_{{\rm Brown}}$ and $p_{{\rm Tallon}}$
for CLBLCO differ slightly from those presented in \ ref. \cite{lt22}, for
two reasons: (a) the calculations in\ our earlier paper did not include the
effect of the displacement of O$_{c}$ and O$_{\alpha }$ from their ``ideal''
positions, and (b) the old values of the oxygen concentration $y$ were taken
from titration, while the present ones were derived from the Rietveld
refinement.

In summary: the BVS method is found to be a very useful tool in improving
the detailed knowledge of the crystal structure. \ In addition, it appears
to provide a good representation of the oxidation states of the CuI and CuII
layers and thus to give the number of Cu$^{{\bf +3}}$\ ions ({\it i.e. }%
holes){\it \ }in the CuII layer. \ But we are forced to conclude,
regretfully, that because the BVS method gives neither the concentration of
mobile charges nor that of Cooper pairs (which can be determined from $\mu $%
sr experiments), it does not contribute to our understanding of the
superconductivity of the cuprates{\bf . \ \ }Although the method is
empirical, and therefore there is no {\it a priori }\ way to determine
whether Brown's or Tallon's method is most reliable, it appears to us that
Brown's method is preferable for the following reasons: \ (a) the better
agreement with the results of Zho and Tafto, (b) the fact (see Fig. 3) that $%
p_{{\rm Brown}}$ \ follows the oxygen concentration $y$ linearly (while $p_{%
{\rm Tallon}}$ does not) appeals to us.

The value of BVS$_{{\rm O}_{\alpha }}$ and the adp $U_{{\rm CuI}}$ are still
slightly anomalous; this is probably a consequence of neglecting the
displacement of CuI from its ideal position when those of its neighboring  O$%
_{\alpha }$ sites which are  occupied are not arranged symmetrically.
However, \ this will not exert any significant influence on the CuII layer.
\ 

\section{\protect\bigskip {\bf Table I}\protect\smallskip}

\begin{tabular}{|l|l|l|l|l|l|l|l|}
\hline
Ion & Cu$^{+}$ & Cu$^{+2}$ & Cu$^{+3}$ & Y$^{+3}$ & Ca$^{+2}$ & La$^{+3}$ & 
Ba$^{+2}$ \\ \hline
$r_{0}$ (\AA ) & 1.600 & 1.679 & 1.730 & 2.019 & 1.967 & 2.172 & 2.285 \\ 
\hline
\end{tabular}

\medskip

\section{Table 2}

\begin{tabular}{cccccc}
\cline{1-1}\cline{3-6}\cline{6-6}
\multicolumn{1}{|c}{Ca conc. $x$} & \multicolumn{1}{|c}{} & 
\multicolumn{1}{|c}{} &  & 0.4 & \multicolumn{1}{c|}{} \\ 
\cline{1-1}\cline{3-6}\cline{6-6}
\multicolumn{1}{|c}{O conc. $y$} & \multicolumn{1}{|c}{} & 
\multicolumn{1}{|c}{} &  & 6.898 & \multicolumn{1}{c|}{} \\ 
\cline{1-1}\cline{3-6}\cline{6-6}
\multicolumn{1}{|c}{$T_{c}$} & \multicolumn{1}{|c}{} & \multicolumn{1}{|c}{}
&  & 33.1 & \multicolumn{1}{c|}{} \\ \cline{1-1}\cline{3-6}\cline{6-6}
&  &  &  &  &  \\ \cline{1-1}\cline{3-6}\cline{6-6}
\multicolumn{1}{|c}{Lattice} & \multicolumn{1}{|c}{} & \multicolumn{1}{|c}{$%
a $} & \multicolumn{1}{|c}{$b$} & \multicolumn{1}{|c|}{$c$} & 
\multicolumn{1}{|c|}{adp.} \\ \cline{3-5}
\multicolumn{1}{|c}{consts.} & \multicolumn{1}{|c}{} & \multicolumn{1}{|c}{
3.879} & \multicolumn{1}{|c}{3.879} & \multicolumn{1}{|c|}{11.702} & 
\multicolumn{1}{|c|}{$(U)$} \\ \cline{1-1}\cline{3-6}\cline{6-6}
&  &  &  &  &  \\ \cline{1-1}\cline{3-6}\cline{6-6}
\multicolumn{1}{|c}{Ion} & \multicolumn{1}{|c}{} & \multicolumn{1}{|c}{$\xi $%
} & \multicolumn{1}{|c}{$\eta $} & \multicolumn{1}{|c|}{$\zeta $} & 
\multicolumn{1}{|c|}{} \\ \cline{1-1}\cline{3-6}\cline{6-6}
\multicolumn{1}{|c}{CuI} & \multicolumn{1}{|c}{} & \multicolumn{1}{|c}{0} & 
\multicolumn{1}{|c}{0} & \multicolumn{1}{|c|}{0} & \multicolumn{1}{|c|}{2.34}
\\ \cline{1-1}\cline{3-6}\cline{6-6}
\multicolumn{1}{|c}{CuII} & \multicolumn{1}{|c}{} & \multicolumn{1}{|c}{0} & 
\multicolumn{1}{|c}{0} & \multicolumn{1}{|c|}{4.121} & \multicolumn{1}{|c|}{
0.63} \\ \cline{1-1}\cline{3-6}\cline{6-6}
\multicolumn{1}{|c}{Ca} & \multicolumn{1}{|c}{} & \multicolumn{1}{|c}{1.940}
& \multicolumn{1}{|c}{1.940} & \multicolumn{1}{|c|}{5.856} & 
\multicolumn{1}{|c|}{0.69} \\ \cline{1-1}\cline{3-6}\cline{6-6}
\multicolumn{1}{|c}{La$_{{\rm (Y)}}$} & \multicolumn{1}{|c}{} & 
\multicolumn{1}{|c}{1.940} & \multicolumn{1}{|c}{1.940} & 
\multicolumn{1}{|c|}{5.856} & \multicolumn{1}{|c|}{0.69} \\ 
\cline{1-1}\cline{3-6}\cline{6-6}
\multicolumn{1}{|c}{Ba} & \multicolumn{1}{|c}{} & \multicolumn{1}{|c}{1.940}
& \multicolumn{1}{|c}{1.940} & \multicolumn{1}{|c|}{2.122} & 
\multicolumn{1}{|c|}{1.04} \\ \cline{1-1}\cline{3-6}\cline{6-6}
\multicolumn{1}{|c}{La$_{{\rm (Ba)}}$} & \multicolumn{1}{|c}{} & 
\multicolumn{1}{|c}{1.940} & \multicolumn{1}{|c}{1.940} & 
\multicolumn{1}{|c|}{2.359} & \multicolumn{1}{|c|}{1.04} \\ 
\cline{1-1}\cline{3-6}\cline{6-6}
\multicolumn{1}{|c}{O$_{\alpha }$} & \multicolumn{1}{|c}{} & 
\multicolumn{1}{|c}{1.940} & \multicolumn{1}{|c}{0} & \multicolumn{1}{|c|}{0}
& \multicolumn{1}{|c|}{7.13} \\ \cline{1-1}\cline{3-6}\cline{6-6}
\multicolumn{1}{|c}{O$_{p}$} & \multicolumn{1}{|c}{} & \multicolumn{1}{|c}{
1.940} & \multicolumn{1}{|c}{0} & \multicolumn{1}{|c|}{4.271} & 
\multicolumn{1}{|c|}{0.74} \\ \cline{1-1}\cline{3-6}\cline{6-6}
\multicolumn{1}{|c}{O$_{c}$} & \multicolumn{1}{|c}{} & \multicolumn{1}{|c}{0}
& \multicolumn{1}{|c}{0} & \multicolumn{1}{|c|}{1.805} & 
\multicolumn{1}{|c|}{1.35} \\ \cline{1-1}\cline{3-6}\cline{6-6}
\end{tabular}

\medskip \eject

\section{\noindent Table 3}

\bigskip ({\bf a)}

\begin{tabular}{lllllll}
\cline{1-1}\cline{3-7}
\multicolumn{1}{|l}{Calcium conc. \ $x$} & \multicolumn{1}{|l}{} & 
\multicolumn{1}{|l}{} &  & 0.1 &  & \multicolumn{1}{l|}{} \\ 
\cline{1-1}\cline{3-7}
&  &  &  &  &  &  \\ \cline{1-1}\cline{3-7}\cline{4-4}\cline{6-6}
\multicolumn{1}{|l}{Oxygen conc. $y$} & \multicolumn{1}{|l}{} & 
\multicolumn{1}{|l}{7.022} & \multicolumn{1}{|l}{7.056} & 
\multicolumn{1}{|l}{7.136} & \multicolumn{1}{|l}{7.232} & 
\multicolumn{1}{|l|}{7.282} \\ \cline{1-1}\cline{3-7}\cline{4-4}\cline{6-6}
\multicolumn{1}{|l}{$T_{c}$} & \multicolumn{1}{|l}{} & \multicolumn{1}{|l}{
30.9} & \multicolumn{1}{|l}{42.6} & \multicolumn{1}{|l}{52.6} & 
\multicolumn{1}{|l}{41.4} & \multicolumn{1}{|l|}{5.0} \\ 
\cline{1-1}\cline{3-7}
\multicolumn{1}{|l}{$r(${\rm CuI to O}$_{\alpha })$} & \multicolumn{1}{|l}{}
& \multicolumn{1}{|l}{1.953} & \multicolumn{1}{|l}{1.952} & 
\multicolumn{1}{|l}{1.953} & \multicolumn{1}{|l}{1.953} & 
\multicolumn{1}{|l|}{1.954} \\ \cline{1-1}\cline{3-7}
\multicolumn{1}{|l}{$r(${\rm CuI to O}$_{c})$} & \multicolumn{1}{|l}{} & 
\multicolumn{1}{|l}{1.838} & \multicolumn{1}{|l}{1.844} & 
\multicolumn{1}{|l}{1.850} & \multicolumn{1}{|l}{1.862} & 
\multicolumn{1}{|l|}{1.864} \\ \cline{1-1}\cline{3-7}
\multicolumn{1}{|l}{$r(${\rm CuII to O}$_{p})$} & \multicolumn{1}{|l}{} & 
\multicolumn{1}{|l}{1.964} & \multicolumn{1}{|l}{1.964} & 
\multicolumn{1}{|l}{1.965} & \multicolumn{1}{|l}{1.964} & 
\multicolumn{1}{|l|}{1.965} \\ \cline{1-1}\cline{3-7}
\multicolumn{1}{|l}{$r(${\rm CuII to O}$_{c})$} & \multicolumn{1}{|l}{} & 
\multicolumn{1}{|l}{2.243} & \multicolumn{1}{|l}{2.228} & 
\multicolumn{1}{|l}{2.215} & \multicolumn{1}{|l}{2.192} & 
\multicolumn{1}{|l|}{2.178} \\ \cline{1-1}\cline{3-7}
\multicolumn{1}{|l}{$r(${\rm Ca to O}$_{p})$} & \multicolumn{1}{|l}{} & 
\multicolumn{1}{|l}{2.518} & \multicolumn{1}{|l}{2.517} & 
\multicolumn{1}{|l}{2.517} & \multicolumn{1}{|l}{2.523} & 
\multicolumn{1}{|l|}{2.535} \\ \cline{1-1}\cline{3-7}
\multicolumn{1}{|l}{$r(${\rm La}$_{({\rm Y}\}}${\rm \ to O}$_{p})$} & 
\multicolumn{1}{|l}{} & \multicolumn{1}{|l}{2.518} & \multicolumn{1}{|l}{
2.517} & \multicolumn{1}{|l}{2.517} & \multicolumn{1}{|l}{2.523} & 
\multicolumn{1}{|l|}{2.535} \\ \cline{1-1}\cline{3-7}
\multicolumn{1}{|l}{$r(${\rm Ba}\ {\rm to O}$_{\alpha })$} & 
\multicolumn{1}{|l}{} & \multicolumn{1}{|l}{2.878} & \multicolumn{1}{|l}{
2.869} & \multicolumn{1}{|l}{2.853} & \multicolumn{1}{|l}{2.847} & 
\multicolumn{1}{|l|}{2.856} \\ \cline{1-1}\cline{3-7}
\multicolumn{1}{|l}{$r(${\rm Ba}\ {\rm to O}$_{p})$} & \multicolumn{1}{|l}{}
& \multicolumn{1}{|l}{2.921} & \multicolumn{1}{|l}{2.928} & 
\multicolumn{1}{|l}{2.943} & \multicolumn{1}{|l}{2.937} & 
\multicolumn{1}{|l|}{2.912} \\ \cline{1-1}\cline{3-7}
\multicolumn{1}{|l}{$r(${\rm Ba}\ {\rm to O}$_{c})$} & \multicolumn{1}{|l}{}
& \multicolumn{1}{|l}{2.776} & \multicolumn{1}{|l}{2.773} & 
\multicolumn{1}{|l}{2.771} & \multicolumn{1}{|l}{2.770} & 
\multicolumn{1}{|l|}{2.773} \\ \cline{1-1}\cline{3-7}
\multicolumn{1}{|l}{$r(${\rm La}$_{({\rm Ba)}}$ {\rm to\ O}$_{\alpha })$} & 
\multicolumn{1}{|l}{} & \multicolumn{1}{|l}{3.019} & \multicolumn{1}{|l}{
2.974} & \multicolumn{1}{|l}{2.977} & \multicolumn{1}{|l}{2.931} & 
\multicolumn{1}{|l|}{2.875} \\ \cline{1-1}\cline{3-7}
\multicolumn{1}{|l}{$r(${\rm La}$_{({\rm Ba)}}$ {\rm to\ O}$_{p})$} & 
\multicolumn{1}{|l}{} & \multicolumn{1}{|l}{2.784} & \multicolumn{1}{|l}{
2.825} & \multicolumn{1}{|l}{2.820} & \multicolumn{1}{|l}{2.853} & 
\multicolumn{1}{|l|}{2.893} \\ \cline{1-1}\cline{3-7}
\multicolumn{1}{|l}{$r(${\rm La}$_{({\rm Ba)}}$ {\rm to\ O}$_{c})$} & 
\multicolumn{1}{|l}{} & \multicolumn{1}{|l}{2.801} & \multicolumn{1}{|l}{
2.790} & \multicolumn{1}{|l}{2.790} & \multicolumn{1}{|l}{2.780} & 
\multicolumn{1}{|l|}{2.775} \\ \cline{1-1}\cline{3-7}
\end{tabular}

\bigskip \eject

({\bf b)}

\smallskip

\begin{tabular}{llllllllll}
\cline{1-1}\cline{3-10}
\multicolumn{1}{|l}{Calcium conc. \ $x$} & \multicolumn{1}{|l}{} & 
\multicolumn{1}{|l}{} &  &  & 0.4 &  &  &  & \multicolumn{1}{l|}{} \\ 
\cline{1-1}\cline{3-10}
&  &  &  &  &  &  &  &  &  \\ \cline{1-1}\cline{3-10}
\multicolumn{1}{|l}{Oxygen conc. $y$} & \multicolumn{1}{|l}{} & 
\multicolumn{1}{|l}{6.852} & \multicolumn{1}{|l}{6.898} & 
\multicolumn{1}{|l}{7.008} & \multicolumn{1}{|l}{7.054} & 
\multicolumn{1}{|l}{7.158} & \multicolumn{1}{|l}{7.176} & 
\multicolumn{1}{|l}{7.244} & \multicolumn{1}{|l|}{7.290} \\ 
\cline{1-1}\cline{3-4}\cline{3-10}\cline{5-10}
\multicolumn{1}{|l}{$T_{c}$} & \multicolumn{1}{|l}{} & \multicolumn{1}{|l}{
12.6} & \multicolumn{1}{|l}{33.1} & \multicolumn{1}{|l}{54.6} & 
\multicolumn{1}{|l}{71.4} & \multicolumn{1}{|l}{79.4} & \multicolumn{1}{|l}{
80.3} & \multicolumn{1}{|l}{75.9} & \multicolumn{1}{|l|}{60.8} \\ 
\cline{1-1}\cline{3-6}\cline{4-4}\cline{7-10}
\multicolumn{1}{|l}{$r(${\rm CuI to O}$_{\alpha })$} & \multicolumn{1}{|l}{}
& \multicolumn{1}{|l}{1.941} & \multicolumn{1}{|l}{1.940} & 
\multicolumn{1}{|l}{1.939} & \multicolumn{1}{|l}{1.938} & 
\multicolumn{1}{|l}{1.938} & \multicolumn{1}{|l}{1.937} & 
\multicolumn{1}{|l}{1.937} & \multicolumn{1}{|l|}{1.938} \\ 
\cline{1-1}\cline{3-10}
\multicolumn{1}{|l}{$r(${\rm CuI to O}$_{c})$} & \multicolumn{1}{|l}{} & 
\multicolumn{1}{|l}{1.805} & \multicolumn{1}{|l}{1.812} & 
\multicolumn{1}{|l}{1.823} & \multicolumn{1}{|l}{1.827} & 
\multicolumn{1}{|l}{1.833} & \multicolumn{1}{|l}{1.838} & 
\multicolumn{1}{|l}{1.844} & \multicolumn{1}{|l|}{1.849} \\ 
\cline{1-1}\cline{3-10}
\multicolumn{1}{|l}{$r(${\rm CuII to O}$_{p})$} & \multicolumn{1}{|l}{} & 
\multicolumn{1}{|l}{1.946} & \multicolumn{1}{|l}{1.946} & 
\multicolumn{1}{|l}{1.946} & \multicolumn{1}{|l}{1.946} & 
\multicolumn{1}{|l}{1.946} & \multicolumn{1}{|l}{1.945} & 
\multicolumn{1}{|l}{1.945} & \multicolumn{1}{|l|}{1.945} \\ 
\cline{1-1}\cline{3-10}
\multicolumn{1}{|l}{$r(${\rm CuII to O}$_{c})$} & \multicolumn{1}{|l}{} & 
\multicolumn{1}{|l}{2.315} & \multicolumn{1}{|l}{2.300} & 
\multicolumn{1}{|l}{2.275} & \multicolumn{1}{|l}{2.261} & 
\multicolumn{1}{|l}{2.251} & \multicolumn{1}{|l}{2.241} & 
\multicolumn{1}{|l}{2.229} & \multicolumn{1}{|l|}{2.225} \\ 
\cline{1-1}\cline{3-10}
\multicolumn{1}{|l}{$r(${\rm Ca to O}$_{p})$} & \multicolumn{1}{|l}{} & 
\multicolumn{1}{|l}{2.505} & \multicolumn{1}{|l}{2.503} & 
\multicolumn{1}{|l}{2.502} & \multicolumn{1}{|l}{2.502} & 
\multicolumn{1}{|l}{2.502} & \multicolumn{1}{|l}{2.501} & 
\multicolumn{1}{|l}{2.502} & \multicolumn{1}{|l|}{2.506} \\ 
\cline{1-1}\cline{3-10}
\multicolumn{1}{|l}{$r(${\rm La}$_{({\rm Y}\}}${\rm \ to O}$_{p})$} & 
\multicolumn{1}{|l}{} & \multicolumn{1}{|l}{2.505} & \multicolumn{1}{|l}{
2.503} & \multicolumn{1}{|l}{2.502} & \multicolumn{1}{|l}{2.502} & 
\multicolumn{1}{|l}{2.502} & \multicolumn{1}{|l}{2.501} & 
\multicolumn{1}{|l}{2.502} & \multicolumn{1}{|l|}{2.506} \\ 
\cline{1-1}\cline{3-10}
\multicolumn{1}{|l}{$r(${\rm Ba}\ {\rm to O}$_{\alpha })$} & 
\multicolumn{1}{|l}{} & \multicolumn{1}{|l}{2.876} & \multicolumn{1}{|l}{
2.870} & \multicolumn{1}{|l}{2.849} & \multicolumn{1}{|l}{2.838} & 
\multicolumn{1}{|l}{2.832} & \multicolumn{1}{|l}{2.823} & 
\multicolumn{1}{|l}{2.823} & \multicolumn{1}{|l|}{2.828} \\ 
\cline{1-1}\cline{3-10}
\multicolumn{1}{|l}{$r(${\rm Ba} {\rm to O}$_{p})$} & \multicolumn{1}{|l}{}
& \multicolumn{1}{|l}{2.895} & \multicolumn{1}{|l}{2.898} & 
\multicolumn{1}{|l}{2.915} & \multicolumn{1}{|l}{2.920} & 
\multicolumn{1}{|l}{2.925} & \multicolumn{1}{|l}{2.928} & 
\multicolumn{1}{|l}{2.926} & \multicolumn{1}{|l|}{2.918} \\ 
\cline{1-1}\cline{3-10}
\multicolumn{1}{|l}{$r(${\rm Ba} {\rm to O}$_{c})$} & \multicolumn{1}{|l}{}
& \multicolumn{1}{|l}{2.763} & \multicolumn{1}{|l}{2.760} & 
\multicolumn{1}{|l}{2.755} & \multicolumn{1}{|l}{2.752} & 
\multicolumn{1}{|l}{2.751} & \multicolumn{1}{|l}{2.748} & 
\multicolumn{1}{|l}{2.748} & \multicolumn{1}{|l|}{2.749} \\ 
\cline{1-1}\cline{3-10}
\multicolumn{1}{|l}{$r(${\rm La}$_{({\rm Ba)}}$ {\rm to\ O}$_{\alpha })$} & 
\multicolumn{1}{|l}{} & \multicolumn{1}{|l}{3.054} & \multicolumn{1}{|l}{
3.055} & \multicolumn{1}{|l}{3.041} & \multicolumn{1}{|l}{3.023} & 
\multicolumn{1}{|l}{3.001} & \multicolumn{1}{|l}{2.998} & 
\multicolumn{1}{|l}{2.979} & \multicolumn{1}{|l|}{2.966} \\ 
\cline{1-1}\cline{3-10}
\multicolumn{1}{|l}{$r(${\rm La}$_{({\rm Ba)}}$ {\rm to O}$_{p})$} & 
\multicolumn{1}{|l}{} & \multicolumn{1}{|l}{2.724} & \multicolumn{1}{|l}{
2.721} & \multicolumn{1}{|l}{2.729} & \multicolumn{1}{|l}{2.740} & 
\multicolumn{1}{|l}{2.759} & \multicolumn{1}{|l}{2.757} & 
\multicolumn{1}{|l}{2.772} & \multicolumn{1}{|l|}{2.782} \\ 
\cline{1-1}\cline{3-10}
\multicolumn{1}{|l}{$r(${\rm La}$_{({\rm Ba)}}$ {\rm to O}$_{c})$} & 
\multicolumn{1}{|l}{} & \multicolumn{1}{|l}{2.800} & \multicolumn{1}{|l}{
2.797} & \multicolumn{1}{|l}{2.791} & \multicolumn{1}{|l}{2.785} & 
\multicolumn{1}{|l}{2.779} & \multicolumn{1}{|l}{2.776} & 
\multicolumn{1}{|l}{2.772} & \multicolumn{1}{|l|}{2.770} \\ 
\cline{1-1}\cline{3-10}
\end{tabular}

\medskip \eject

\section{Table 4}

{\bf (a)}

\begin{tabular}{lllllll}
\cline{1-1}\cline{3-7}
\multicolumn{1}{|l}{Ca conc. $x$} & \multicolumn{1}{|l}{} & 
\multicolumn{1}{|l}{} &  & 0.1 &  & \multicolumn{1}{l|}{} \\ 
\cline{1-1}\cline{3-7}
&  &  &  &  &  &  \\ \cline{1-1}\cline{3-7}
\multicolumn{1}{|l}{O conc. $y$} & \multicolumn{1}{|l}{} & 
\multicolumn{1}{|l}{7.022} & \multicolumn{1}{|l}{7.056} & 
\multicolumn{1}{|l}{7.136} & \multicolumn{1}{|l}{7.232} & 
\multicolumn{1}{|l|}{7.282} \\ \cline{1-1}\cline{3-7}
\multicolumn{1}{|l}{BVS$_{{\rm Ca}}^{0}$} & \multicolumn{1}{|l}{} & 
\multicolumn{1}{|l}{1.802} & \multicolumn{1}{|l}{1.810} & 
\multicolumn{1}{|l}{1.808} & \multicolumn{1}{|l}{1.782} & 
\multicolumn{1}{|l|}{1.721} \\ \cline{1-1}\cline{3-7}
\multicolumn{1}{|l}{BVS$_{{\rm La(Y)}}^{0}$} & \multicolumn{1}{|l}{} & 
\multicolumn{1}{|l}{3.137} & \multicolumn{1}{|l}{3.150} & 
\multicolumn{1}{|l}{3.147} & \multicolumn{1}{|l}{3.101} & 
\multicolumn{1}{|l|}{2.995} \\ \cline{1-1}\cline{3-7}
\multicolumn{1}{|l}{BVS$_{{\rm La(Ba)}}^{0}$} & \multicolumn{1}{|l}{} & 
\multicolumn{1}{|l}{1.702} & \multicolumn{1}{|l}{1.680} & 
\multicolumn{1}{|l}{1.704} & \multicolumn{1}{|l}{1.724} & 
\multicolumn{1}{|l|}{1.738} \\ \cline{1-1}\cline{3-7}
\multicolumn{1}{|l}{BVS$_{{\rm Ba}}^{0}$} & \multicolumn{1}{|l}{} & 
\multicolumn{1}{|l}{2.188} & \multicolumn{1}{|l}{2.208} & 
\multicolumn{1}{|l}{2.239} & \multicolumn{1}{|l}{2.306} & 
\multicolumn{1}{|l|}{2.354} \\ \cline{1-1}\cline{3-7}
\multicolumn{1}{|l}{BVS$_{{\rm Op}}^{0}$} & \multicolumn{1}{|l}{} & 
\multicolumn{1}{|l}{2.052} & \multicolumn{1}{|l}{2.045} & 
\multicolumn{1}{|l}{2.032} & \multicolumn{1}{|l}{2.027} & 
\multicolumn{1}{|l|}{2.016} \\ \cline{1-1}\cline{3-7}
BVS$_{{\rm O\alpha }}^{0}$ & \multicolumn{1}{|l}{} & 1.747 & 
\multicolumn{1}{|l}{1.777} & \multicolumn{1}{|l}{1.818} & 
\multicolumn{1}{|l}{1.854} & \multicolumn{1}{|l|}{1.854} \\ 
\cline{1-1}\cline{3-7}
\multicolumn{1}{|l}{BVS$_{{\rm Oc}}^{0}$} & \multicolumn{1}{|l}{} & 
\multicolumn{1}{|l}{1.916} & \multicolumn{1}{|l}{1.926} & 
\multicolumn{1}{|l}{1.937} & \multicolumn{1}{|l}{1.946} & 
\multicolumn{1}{|l|}{1.954} \\ \cline{1-1}\cline{3-7}
\end{tabular}

{\bf (b)}

\begin{tabular}{llllllllll}
\cline{1-1}\cline{3-10}\cline{6-6}
\multicolumn{1}{|l}{Ca conc. $x$} & \multicolumn{1}{|l}{} & 
\multicolumn{1}{|l}{} &  &  &  & 0.4 &  &  & \multicolumn{1}{l|}{} \\ 
\cline{1-1}\cline{3-10}\cline{6-6}\cline{10-10}
&  &  &  &  &  &  &  &  &  \\ \cline{1-1}\cline{3-10}\cline{10-10}
\multicolumn{1}{|l}{O conc. $y$} & \multicolumn{1}{|l}{} & 
\multicolumn{1}{|l}{6.852} & \multicolumn{1}{|l}{6.898} & 
\multicolumn{1}{|l}{7.008} & \multicolumn{1}{|l}{7.054} & 
\multicolumn{1}{|l}{7.158} & \multicolumn{1}{|l}{7.176} & 
\multicolumn{1}{|l}{7.244} & \multicolumn{1}{|l|}{7.290} \\ 
\cline{1-1}\cline{3-10}
\multicolumn{1}{|l}{BVS$_{{\rm Ca}}^{0}$} & \multicolumn{1}{|l}{} & 
\multicolumn{1}{|l}{1.866} & \multicolumn{1}{|l}{1.878} & 
\multicolumn{1}{|l}{1.882} & \multicolumn{1}{|l}{1.883} & 
\multicolumn{1}{|l}{1.886} & \multicolumn{1}{|l}{1.890} & 
\multicolumn{1}{|l}{1.884} & \multicolumn{1}{|l|}{1.862} \\ 
\cline{1-1}\cline{3-10}
\multicolumn{1}{|l}{BVS$_{{\rm La(Y)}}^{0}$} & \multicolumn{1}{|l}{} & 
\multicolumn{1}{|l}{3.248} & \multicolumn{1}{|l}{3.268} & 
\multicolumn{1}{|l}{3.275} & \multicolumn{1}{|l}{3.277} & 
\multicolumn{1}{|l}{3.283} & \multicolumn{1}{|l}{3.289} & 
\multicolumn{1}{|l}{3.279} & \multicolumn{1}{|l|}{3.241} \\ 
\cline{1-1}\cline{3-10}
\multicolumn{1}{|l}{BVS$_{{\rm Ba}}^{0}$} & \multicolumn{1}{|l}{} & 
\multicolumn{1}{|l}{2.214} & \multicolumn{1}{|l}{2.242} & 
\multicolumn{1}{|l}{2.291} & \multicolumn{1}{|l}{2.324} & 
\multicolumn{1}{|l}{2.374} & \multicolumn{1}{|l}{2.397} & 
\multicolumn{1}{|l}{2.433} & \multicolumn{1}{|l|}{2.458} \\ 
\cline{1-1}\cline{3-10}
\multicolumn{1}{|l}{BVS$_{{\rm La(Ba)}}^{0}$} & \multicolumn{1}{|l}{} & 
\multicolumn{1}{|l}{1.789} & \multicolumn{1}{|l}{1.811} & 
\multicolumn{1}{|l}{1.830} & \multicolumn{1}{|l}{1.836} & 
\multicolumn{1}{|l}{1.840} & \multicolumn{1}{|l}{1.857} & 
\multicolumn{1}{|l}{1.861} & \multicolumn{1}{|l|}{1.867} \\ 
\cline{1-1}\cline{3-10}
\multicolumn{1}{|l}{BVS$_{{\rm Op}}^{0}$} & \multicolumn{1}{|l}{} & 
\multicolumn{1}{|l}{2.076} & \multicolumn{1}{|l}{2.083} & 
\multicolumn{1}{|l}{2.071} & \multicolumn{1}{|l}{2.068} & 
\multicolumn{1}{|l}{2.060} & \multicolumn{1}{|l}{2.065} & 
\multicolumn{1}{|l}{2.059} & \multicolumn{1}{|l|}{2.052} \\ 
\cline{1-1}\cline{3-10}
\multicolumn{1}{|l}{BVS$_{{\rm O\alpha }}^{0}$} & \multicolumn{1}{|l}{} & 
\multicolumn{1}{|l}{1.710} & \multicolumn{1}{|l}{1.726} & 
\multicolumn{1}{|l}{1.782} & \multicolumn{1}{|l}{1.815} & 
\multicolumn{1}{|l}{1.855} & \multicolumn{1}{|l}{1.874} & 
\multicolumn{1}{|l}{1.894} & \multicolumn{1}{|l|}{1.895} \\ 
\cline{1-1}\cline{3-10}
\multicolumn{1}{|l}{BVS$_{\text{{\rm O}{\it c}}}^{0}$} & \multicolumn{1}{|l}{
} & \multicolumn{1}{|l}{1.917} & \multicolumn{1}{|l}{1.922} & 
\multicolumn{1}{|l}{1.939} & \multicolumn{1}{|l}{1.954} & 
\multicolumn{1}{|l}{1.971} & \multicolumn{1}{|l}{1.975} & 
\multicolumn{1}{|l}{1.984} & \multicolumn{1}{|l|}{1.980} \\ 
\cline{1-1}\cline{3-10}
\end{tabular}

\mbox{$\backslash$}%

\medskip

\section{Table 5}

\begin{tabular}{cccccc}
\cline{1-1}\cline{3-6}\cline{6-6}
\multicolumn{1}{|c}{Ca conc. $x$} & \multicolumn{1}{|c}{} & 
\multicolumn{1}{|c}{} &  & 0.4 & \multicolumn{1}{c|}{} \\ 
\cline{1-1}\cline{3-6}\cline{6-6}
\multicolumn{1}{|c}{O conc. $y$} & \multicolumn{1}{|c}{} & 
\multicolumn{1}{|c}{} &  & 6.924 & \multicolumn{1}{c|}{} \\ 
\cline{1-1}\cline{3-6}\cline{6-6}
\multicolumn{1}{|c}{$T_{c}$} & \multicolumn{1}{|c}{} & \multicolumn{1}{|c}{}
&  & 33.1 & \multicolumn{1}{c|}{} \\ \cline{1-1}\cline{3-6}
&  &  &  &  &  \\ \cline{1-1}\cline{3-6}\cline{6-6}
\multicolumn{1}{|c}{Lattice} & \multicolumn{1}{|c}{} & \multicolumn{1}{|c}{$%
a $} & \multicolumn{1}{|c}{$b$} & \multicolumn{1}{|c|}{$c$} & 
\multicolumn{1}{|c|}{adp} \\ \cline{3-5}
\multicolumn{1}{|c}{consts.} & \multicolumn{1}{|c}{} & \multicolumn{1}{|c}{
3.879} & \multicolumn{1}{|c}{3.879} & \multicolumn{1}{|c|}{11.701} & 
\multicolumn{1}{|c|}{$U$(\AA $^{2}$)} \\ \cline{1-1}\cline{3-6}\cline{6-6}
&  &  &  &  &  \\ \cline{1-1}\cline{3-6}\cline{6-6}
\multicolumn{1}{|c}{Ion} & \multicolumn{1}{|c}{} & \multicolumn{1}{|c}{$\xi $%
} & \multicolumn{1}{|c}{$\eta $} & \multicolumn{1}{|c|}{$\zeta $} & 
\multicolumn{1}{|c|}{} \\ \cline{1-1}\cline{3-6}\cline{6-6}
\multicolumn{1}{|c}{CuI} & \multicolumn{1}{|c}{} & \multicolumn{1}{|c}{0} & 
\multicolumn{1}{|c}{0} & \multicolumn{1}{|c|}{0} & \multicolumn{1}{|c|}{2.39}
\\ \cline{1-1}\cline{3-6}\cline{6-6}
\multicolumn{1}{|c}{CuII} & \multicolumn{1}{|c}{} & \multicolumn{1}{|c}{0} & 
\multicolumn{1}{|c}{0} & \multicolumn{1}{|c|}{4.110} & \multicolumn{1}{|c|}{
0.58} \\ \cline{1-1}\cline{3-6}\cline{6-6}
\multicolumn{1}{|c}{Ca} & \multicolumn{1}{|c}{} & \multicolumn{1}{|c}{1.940}
& \multicolumn{1}{|c}{1.940} & \multicolumn{1}{|c|}{5.851} & 
\multicolumn{1}{|c|}{0.66} \\ \cline{1-1}\cline{3-6}\cline{6-6}
\multicolumn{1}{|c}{La$_{{\rm (Y)}}$} & \multicolumn{1}{|c}{} & 
\multicolumn{1}{|c}{1.940} & \multicolumn{1}{|c}{1.940} & 
\multicolumn{1}{|c|}{5.851} & \multicolumn{1}{|c|}{0.66} \\ 
\cline{1-1}\cline{3-6}\cline{6-6}
\multicolumn{1}{|c}{Ba} & \multicolumn{1}{|c}{} & \multicolumn{1}{|c}{1.940}
& \multicolumn{1}{|c}{1.940} & \multicolumn{1}{|c|}{2.334} & 
\multicolumn{1}{|c|}{0.90} \\ \cline{1-1}\cline{3-6}\cline{6-6}
\multicolumn{1}{|c}{La$_{{\rm (Ba)}}$} & \multicolumn{1}{|c}{} & 
\multicolumn{1}{|c}{1.940} & \multicolumn{1}{|c}{1.940} & 
\multicolumn{1}{|c|}{2.062} & \multicolumn{1}{|c|}{0.90} \\ 
\cline{1-1}\cline{3-6}\cline{6-6}
\multicolumn{1}{|c}{O$_{\alpha }(1)$} & \multicolumn{1}{|c}{} & 
\multicolumn{1}{|c}{1.940} & \multicolumn{1}{|c}{0} & \multicolumn{1}{|c|}{0}
& \multicolumn{1}{|c|}{1.80} \\ \cline{1-1}\cline{3-6}\cline{6-6}
\multicolumn{1}{|c}{O$_{\alpha }(2)$} & \multicolumn{1}{|c}{} & 
\multicolumn{1}{|c}{1.940} & \multicolumn{1}{|c}{0.755} & 
\multicolumn{1}{|c|}{0} & \multicolumn{1}{|c|}{1.80} \\ 
\cline{1-1}\cline{3-6}\cline{6-6}
\multicolumn{1}{|c}{O$_{p}$} & \multicolumn{1}{|c}{} & \multicolumn{1}{|c}{
1.940} & \multicolumn{1}{|c}{0} & \multicolumn{1}{|c|}{4.267} & 
\multicolumn{1}{|c|}{0.76} \\ \cline{1-1}\cline{3-6}\cline{6-6}
\multicolumn{1}{|c}{O$_{c}(1)$} & \multicolumn{1}{|c}{} & 
\multicolumn{1}{|c}{0} & \multicolumn{1}{|c}{0} & \multicolumn{1}{|c|}{1.818}
& \multicolumn{1}{|c|}{0.73} \\ \cline{1-1}\cline{3-3}\cline{3-6}\cline{5-6}
\multicolumn{1}{|c}{O$_{c}(2)$} & \multicolumn{1}{|c}{} & 
\multicolumn{1}{|c}{0.266} & \multicolumn{1}{|c}{0.266} & 
\multicolumn{1}{|c|}{1.818} & \multicolumn{1}{|c|}{0.73} \\ 
\cline{1-1}\cline{3-3}\cline{3-6}\cline{5-6}
\multicolumn{1}{|c}{O$_{c}(3)$} & \multicolumn{1}{|c}{} & 
\multicolumn{1}{|c}{0.451} & \multicolumn{1}{|c}{0} & \multicolumn{1}{|c|}{
1.818} & \multicolumn{1}{|c|}{0.73} \\ \cline{1-1}\cline{3-6}\cline{6-6}
\end{tabular}

\section{\protect\large Table 6: \ }

{\large Calcium concentration: (a) x = 0.1,\ (b) x = 0.4, }

{\large next two pages}

\medskip

\bigskip \null\vskip10pt

\vskip10pt

\begin{tabular}{l|lllllll}
\cline{2-2}\cline{4-8}
{\bf (a)} & Oxygen conc. $y$ & \multicolumn{1}{|l}{} & \multicolumn{1}{|l}{
7.038} & 7.084 & 7.158 & 7.180 & \multicolumn{1}{l|}{7.258} \\ 
\cline{2-2}\cline{4-8}
& $r(${\rm CuI to O}$_{\alpha (1)})$ & \multicolumn{1}{|l}{} & 
\multicolumn{1}{|l}{1.953} & \multicolumn{1}{|l}{1.952} & 
\multicolumn{1}{|l}{1.953} & \multicolumn{1}{|l}{1.953} & 
\multicolumn{1}{|l|}{1.954} \\ \cline{2-2}\cline{4-8}
& $r(${\rm CuI to O}$_{\alpha (2)})$ & \multicolumn{1}{|l}{} & 
\multicolumn{1}{|l}{2.024} & \multicolumn{1}{|l}{2.028} & 
\multicolumn{1}{|l}{2.015} & \multicolumn{1}{|l}{2.000} & 
\multicolumn{1}{|l|}{2.001} \\ \cline{2-2}\cline{4-8}
& $r(${\rm CuI to O}$_{c(1)})$ & \multicolumn{1}{|l}{} & \multicolumn{1}{|l}{
1.869} & \multicolumn{1}{|l}{1.868} & \multicolumn{1}{|l}{1.867} & 
\multicolumn{1}{|l}{1.869} & \multicolumn{1}{|l|}{1.869} \\ 
\cline{2-2}\cline{4-8}
& $r(${\rm CuI to O}$_{c(2)})$ & \multicolumn{1}{|l}{} & \multicolumn{1}{|l}{
1.902} & \multicolumn{1}{|l}{1.901} & \multicolumn{1}{|l}{1.898} & 
\multicolumn{1}{|l}{1.898} & \multicolumn{1}{|l|}{1.893} \\ 
\cline{2-2}\cline{4-8}
& $r(${\rm CuI to O}$_{c(3)})$ & \multicolumn{1}{|l}{} & \multicolumn{1}{|l}{
1.920} & \multicolumn{1}{|l}{1.917} & \multicolumn{1}{|l}{1.926} & 
\multicolumn{1}{|l}{1.927} & \multicolumn{1}{|l|}{1.936} \\ 
\cline{2-2}\cline{4-8}
& $r(${\rm CuII to O}$_{p})$ & \multicolumn{1}{|l}{} & \multicolumn{1}{|l}{
1.964} & \multicolumn{1}{|l}{1.964} & \multicolumn{1}{|l}{1.965} & 
\multicolumn{1}{|l}{1.964} & \multicolumn{1}{|l|}{1.964} \\ 
\cline{2-2}\cline{4-8}
& $r(${\rm CuII to O}$_{c(1)})$ & \multicolumn{1}{|l}{} & 
\multicolumn{1}{|l}{2.210} & \multicolumn{1}{|l}{2.203} & 
\multicolumn{1}{|l}{2.197} & \multicolumn{1}{|l}{2.184} & 
\multicolumn{1}{|l|}{2.172} \\ \cline{2-2}\cline{4-8}
& $r(${\rm CuII to O}$_{c(2)})$ & \multicolumn{1}{|l}{} & 
\multicolumn{1}{|l}{2.238} & \multicolumn{1}{|l}{2.232} & 
\multicolumn{1}{|l}{2.223} & \multicolumn{1}{|l}{2.208} & 
\multicolumn{1}{|l|}{2.193} \\ \cline{2-2}\cline{4-8}
& $r(${\rm CuII to O}$_{c(3)})$ & \multicolumn{1}{|l}{} & 
\multicolumn{1}{|l}{2.253} & \multicolumn{1}{|l}{2.246} & 
\multicolumn{1}{|l}{2.247} & \multicolumn{1}{|l}{2.234} & 
\multicolumn{1}{|l|}{2.230} \\ \cline{2-2}\cline{4-8}
& $r(${\rm Ca to O}$_{p})$ & \multicolumn{1}{|l}{} & \multicolumn{1}{|l}{
2.518} & \multicolumn{1}{|l}{2.517} & \multicolumn{1}{|l}{2.518} & 
\multicolumn{1}{|l}{2.524} & \multicolumn{1}{|l|}{2.537} \\ 
\cline{2-2}\cline{4-8}
& $r(${\rm La}$_{({\rm Y}\}}${\rm \ to O}$_{p})$ & \multicolumn{1}{|l}{} & 
\multicolumn{1}{|l}{2.518} & \multicolumn{1}{|l}{2.517} & 
\multicolumn{1}{|l}{2.518} & \multicolumn{1}{|l}{2.524} & 
\multicolumn{1}{|l|}{2.537} \\ \cline{2-2}\cline{4-8}
& $r(${\rm Ba}\ {\rm to O}$_{\alpha (1)})$ & \multicolumn{1}{|l}{} & 
\multicolumn{1}{|l}{2.934} & \multicolumn{1}{|l}{2.921} & 
\multicolumn{1}{|l}{2.913} & \multicolumn{1}{|l}{2.892} & 
\multicolumn{1}{|l|}{2.865} \\ \cline{2-2}\cline{4-8}
& $r(${\rm Ba}\ {\rm to O}$_{\alpha (2)})$ & \multicolumn{1}{|l}{} & 
\multicolumn{1}{|l}{3.312} & \multicolumn{1}{|l}{3.312} & 
\multicolumn{1}{|l}{3.267} & \multicolumn{1}{|l}{3.201} & 
\multicolumn{1}{|l|}{3.174} \\ \cline{2-2}\cline{4-8}
& $r(${\rm Ba}\ {\rm to O}$_{p})$ & \multicolumn{1}{|l}{} & 
\multicolumn{1}{|l}{2.866} & \multicolumn{1}{|l}{2.877} & 
\multicolumn{1}{|l}{2.882} & \multicolumn{1}{|l}{2.891} & 
\multicolumn{1}{|l|}{2.901} \\ \cline{2-2}\cline{4-8}
& $r(${\rm Ba}\ {\rm to O}$_{c(1)})$ & \multicolumn{1}{|l}{} & 
\multicolumn{1}{|l}{2.781} & \multicolumn{1}{|l}{2.778} & 
\multicolumn{1}{|l}{2.778} & \multicolumn{1}{|l}{2.774} & 
\multicolumn{1}{|l|}{2.773} \\ \cline{2-2}\cline{4-8}
& $r_{_{_{\parallel }}}(${\rm Ba}\ {\rm to O}$_{c(2)})$ & 
\multicolumn{1}{|l}{} & \multicolumn{1}{|l}{3.132} & \multicolumn{1}{|l}{
3.132} & \multicolumn{1}{|l}{3.117} & \multicolumn{1}{|l}{3.101} & 
\multicolumn{1}{|l|}{3.074} \\ \cline{2-2}\cline{4-8}
& $r_{\perp }(${\rm Ba}\ {\rm to O}$_{c(2)})$ & \multicolumn{1}{|l}{} & 
\multicolumn{1}{|l}{2.803} & \multicolumn{1}{|l}{2.801} & 
\multicolumn{1}{|l}{2.798} & \multicolumn{1}{|l}{2.793} & 
\multicolumn{1}{|l|}{2.790} \\ \cline{2-2}\cline{4-8}
& $r(${\rm Ba}\ {\rm to O}$_{c(3)})$ & \multicolumn{1}{|l}{} & 
\multicolumn{1}{|l}{3.104} & \multicolumn{1}{|l}{3.098} & 
\multicolumn{1}{|l}{3.128} & \multicolumn{1}{|l}{3.123} & 
\multicolumn{1}{|l|}{3.149} \\ \cline{2-2}\cline{4-8}
& $r(${\rm La}$_{({\rm Ba)}}$ {\rm to\ O}$_{\alpha (1)})$ & 
\multicolumn{1}{|l}{} & \multicolumn{1}{|l}{2.807} & \multicolumn{1}{|l}{
2.812} & \multicolumn{1}{|l}{2.784} & \multicolumn{1}{|l}{2.784} & 
\multicolumn{1}{|l|}{2.840} \\ \cline{2-2}\cline{4-8}
& $r(${\rm La}$_{({\rm Ba)}}$ {\rm to\ O}$_{\alpha (2)})$ & 
\multicolumn{1}{|l}{} & \multicolumn{1}{|l}{2.467} & \multicolumn{1}{|l}{
2.464} & \multicolumn{1}{|l}{2.461} & \multicolumn{1}{|l}{2.499} & 
\multicolumn{1}{|l|}{2.563} \\ \cline{2-2}\cline{4-8}
& $r(${\rm La}$_{({\rm Ba)}}$ {\rm to\ O}$_{p})$ & \multicolumn{1}{|l}{} & 
\multicolumn{1}{|l}{2.995} & \multicolumn{1}{|l}{2.988} & 
\multicolumn{1}{|l}{3.015} & \multicolumn{1}{|l}{3.002} & 
\multicolumn{1}{|l|}{2.926} \\ \cline{2-2}\cline{4-8}
& $r(${\rm La}$_{({\rm Ba)}}$ {\rm to\ O}$_{c(1)})$ & \multicolumn{1}{|l}{}
& \multicolumn{1}{|l}{2.766} & \multicolumn{1}{|l}{2.766} & 
\multicolumn{1}{|l}{2.764} & \multicolumn{1}{|l}{2.764} & 
\multicolumn{1}{|l|}{2.771} \\ \cline{2-2}\cline{4-8}
& $r_{\parallel }(${\rm La}$_{({\rm Ba)}}$ {\rm to\ O}$_{c(2)})$ & 
\multicolumn{1}{|l}{} & \multicolumn{1}{|l}{2.414} & \multicolumn{1}{|l}{
2.410} & \multicolumn{1}{|l}{2.423} & \multicolumn{1}{|l}{2.436} & 
\multicolumn{1}{|l|}{2.470} \\ \cline{2-2}\cline{4-8}
& $r_{\perp }(${\rm La}$_{({\rm Ba)}}$ {\rm to\ O}$_{c(2)})$ & 
\multicolumn{1}{|l}{} & \multicolumn{1}{|l}{2.793} & \multicolumn{1}{|l}{
2.793} & \multicolumn{1}{|l}{2.789} & \multicolumn{1}{|l}{2.786} & 
\multicolumn{1}{|l|}{2.788} \\ \cline{2-2}\cline{4-8}
& $r(${\rm La}$_{({\rm Ba)}}$ {\rm to\ O}$_{c(3)})$ & \multicolumn{1}{|l}{}
& \multicolumn{1}{|l}{2.476} & \multicolumn{1}{|l}{2.479} & 
\multicolumn{1}{|l}{2.453} & \multicolumn{1}{|l}{2.454} & 
\multicolumn{1}{|l|}{2.442} \\ \cline{2-2}\cline{4-8}
& occupancy $n(${\rm O}$_{\alpha (1)})$ & \multicolumn{1}{|l}{} & 
\multicolumn{1}{|l}{0.638} & \multicolumn{1}{|l}{0.660} & 
\multicolumn{1}{|l}{0.666} & \multicolumn{1}{|l}{0.648} & 
\multicolumn{1}{|l|}{0.666} \\ \cline{2-2}\cline{4-8}
& occupancy $n(${\rm O}$_{\alpha (2)})$ & \multicolumn{1}{|l}{} & 
\multicolumn{1}{|l}{0.400} & \multicolumn{1}{|l}{0.424} & 
\multicolumn{1}{|l}{0.492} & \multicolumn{1}{|l}{0.532} & 
\multicolumn{1}{|l|}{0.592} \\ \cline{2-2}\cline{4-8}
\end{tabular}

\begin{tabular}{l|l|l|l|l|l|l|l|l|l|l|}
\cline{2-2}\cline{4-11}
{\bf (b)} & Oxygen conc. $y$ &  & 6.884 & 6.926 & 7.008 & 7.068 & 7.142 & 
7.174 & 7.208 & 7.240 \\ \cline{2-2}\cline{4-11}
& $r(${\rm CuI to O}$_{\alpha (1)})$ &  & 1.940 & 1.940 & 1.939 & 1.938 & 
1.938 & 1.937 & 1.937 & 1.938 \\ \cline{2-2}\cline{4-11}
& $r(${\rm CuI to O}$_{\alpha (2)})$ &  & 2.082 & 2.081 & 2.040 & 2.030 & 
2.027 & 2.181 & 2.015 & 2.012 \\ \cline{2-2}\cline{4-11}
& $r(${\rm CuI to O}$_{c(1)})$ &  & 1.813 & 1.818 & 1.829 & 1.832 & 1.838 & 
1.842 & 1.847 & 1.852 \\ \cline{2-2}\cline{4-11}
& $r(${\rm CuI to O}$_{c(2)})$ &  & 1.853 & 1.857 & 1.863 & 1.865 & 1.869 & 
1.871 & 1.875 & 1.883 \\ \cline{2-2}\cline{4-11}
& $r(${\rm CuI to O}$_{c(3)})$ &  & 1.863 & 1.873 & 1.876 & 1.875 & 1.880 & 
1.896 & 1.893 & 1.897 \\ \cline{2-2}\cline{4-11}
& $r(${\rm CuII to O}$_{p})$ &  & 1.946 & 1.946 & 1.946 & 1.946 & 1.946 & 
1.945 & 1.945 & 1.945 \\ \cline{2-2}\cline{4-11}
& $r(${\rm CuII to O}$_{c(1)})$ &  & 2.305 & 2.292 & 2.266 & 2.254 & 2.243 & 
2.235 & 2.224 & 2.219 \\ \cline{2-2}\cline{4-11}
& $r(${\rm CuII to O}$_{c(2)})$ &  & 2.337 & 2.322 & 2.294 & 2.280 & 2.269 & 
2.259 & 2.247 & 2.245 \\ \cline{2-2}\cline{4-11}
& $r(${\rm CuII to O}$_{c(3)})$ &  & 2.345 & 2.336 & 2.304 & 2.289 & 2.278 & 
2.279 & 2.262 & 2.257 \\ \cline{2-2}\cline{4-11}
& $r(${\rm Ca to O}$_{p})$ &  & 2.506 & 2.504 & 2.502 & 2.503 & 2.502 & 2.502
& 2.504 & 2.508 \\ \cline{2-2}\cline{4-11}
& $r(${\rm La}$_{({\rm Y}\}}${\rm \ to O}$_{p})$ &  & 2.506 & 2.504 & 2.502
& 2.503 & 2.502 & 2.502 & 2.504 & 2.508 \\ \cline{2-2}\cline{4-11}
& $r(${\rm Ba}\ {\rm to O}$_{\alpha (1)})$ &  & 3.041 & 3.035 & 3.019 & 2.998
& 2.984 & 2.979 & 2.963 & 2.952 \\ \cline{2-2}\cline{4-11}
& $r(${\rm Ba}\ {\rm to O}$_{\alpha (2)})$ &  & 3.571 & 3.565 & 3.460 & 3.421
& 3.400 & 3.375 & 3.351 & 3.332 \\ \cline{2-2}\cline{4-11}
& $r(${\rm Ba} {\rm to O}$_{p})$ &  & 2.736 & 2.739 & 2.749 & 2.762 & 2.774
& 2.773 & 2.786 & 2.793 \\ \cline{2-2}\cline{4-11}
& $r(${\rm Ba} {\rm to O}$_{c(1)})$ &  & 2.795 & 2.791 & 2.785 & 2.779 & 
2.775 & 2.772 & 2.768 & 2.767 \\ \cline{2-2}\cline{4-11}
& $r_{_{\parallel }}(${\rm Ba} {\rm to O}$_{c(2)})$ &  & 3.171 & 3.161 & 
3.135 & 3.120 & 3110 & 3.097 & 3.088 & 3.103 \\ \cline{2-2}\cline{4-11}
& $r_{\perp }(${\rm Ba} {\rm to O}$_{c(2)})$ &  & 2.821 & 2.816 & 2.807 & 
2.800 & 2.795 & 2.791 & 2.787 & 2.787 \\ \cline{2-2}\cline{4-11}
& $r(${\rm Ba} {\rm to O}$_{c(3)})$ &  & 3.108 & 3.121 & 3.090 & 3.069 & 
3.066 & 3.102 & 3.074 & 3.068 \\ \cline{2-2}\cline{4-11}
& $r(${\rm La}$_{({\rm Ba)}}$ {\rm to\ O}$_{\alpha (1)})$ &  & 2.840 & 2.831
& 2.810 & 2.799 & 2.797 & 2.788 & 2.789 & 2.794 \\ \cline{2-2}\cline{4-11}
& $r(${\rm La}$_{({\rm Ba)}}$ {\rm to\ O}$_{\alpha (2)})$ &  & 2.389 & 2.378
& 2.417 & 2.420 & 2.424 & 2.429 & 2.438 & 2.451 \\ \cline{2-2}\cline{4-11}
& $r(${\rm La}$_{({\rm Ba)}}$ {\rm to O}$_{p})$ &  & 2.930 & 2.937 & 2,954 & 
2.959 & 2.960 & 2.963 & 2.960 & 2.951 \\ \cline{2-2}\cline{4-11}
& $r(${\rm La}$_{({\rm Ba)}}$ {\rm to O}$_{c(1)})$ &  & 2.757 & 2.754 & 2.750
& 2.748 & 2.747 & 2.745 & 2.744 & 2.746 \\ \cline{2-2}\cline{4-11}
& $r_{\parallel }(${\rm La}$_{({\rm Ba)}}$ {\rm to O}$_{c(2)})$ &  & 2.376 & 
2.380 & 2,396 & 2.403 & 2.409 & 2.417 & 2.422 & 2.407 \\ 
\cline{2-2}\cline{4-11}
& $r_{\perp }(${\rm La}$_{({\rm Ba)}}$ {\rm to O}$_{c(2)})$ &  & 2.783 & 
2.779 & 2.773 & 2.769 & 2.768 & 2.764 & 2.763 & 2.767 \\ 
\cline{2-2}\cline{4-11}
& $r(${\rm La}$_{({\rm Ba)}}$ {\rm to O}$_{c(3)})$ &  & 2.473 & 2.457 & 2.473
& 2.484 & 2.483 & 2.449 & 2.460 & 2.473 \\ \cline{2-2}\cline{4-11}
& occupancy $n(${\rm O}$_{\alpha (1)})$ &  & 0.480 & 0.516 & 0.514 & 0.542 & 
0.566 & 0.578 & 0.596 & 0.582 \\ \cline{2-2}\cline{4-11}
& occupancy $n(${\rm O}$_{\alpha (2)})$ &  & 0.404 & 0.408 & 0.488 & 0.520 & 
0.572 & 0.596 & 0.608 & 0.656 \\ \cline{2-2}\cline{4-11}
\end{tabular}

\noindent

\section{Table 7}

{\bf (a)}

\begin{tabular}{lllllll}
\cline{1-1}\cline{3-7}\cline{7-7}
\multicolumn{1}{|l}{Ca conc. $x$} & \multicolumn{1}{|l}{} & 
\multicolumn{1}{|l}{} &  & 0.1 &  & \multicolumn{1}{|l|}{} \\ 
\cline{1-1}\cline{3-7}\cline{7-7}
&  &  &  &  &  &  \\ \cline{1-1}\cline{3-7}\cline{7-7}
\multicolumn{1}{|l}{O conc. $y$} & \multicolumn{1}{|l}{} & 
\multicolumn{1}{|l}{7.038} & \multicolumn{1}{|l}{7.084} & 
\multicolumn{1}{|l}{7.158} & \multicolumn{1}{|l}{7.180} & 
\multicolumn{1}{|l|}{7.258} \\ \cline{1-1}\cline{3-7}\cline{7-7}
\multicolumn{1}{|l}{BVS$_{{\rm Ca}}$} & \multicolumn{1}{|l}{} & 
\multicolumn{1}{|l}{1.803} & \multicolumn{1}{|l}{1.810} & 
\multicolumn{1}{|l}{1.806} & \multicolumn{1}{|l}{1.777} & 
\multicolumn{1}{|l|}{1.713} \\ \cline{1-1}\cline{3-7}\cline{7-7}
\multicolumn{1}{|l}{BVS$_{{\rm La(Y)}}$} & \multicolumn{1}{|l}{} & 
\multicolumn{1}{|l}{3.138} & \multicolumn{1}{|l}{3.150} & 
\multicolumn{1}{|l}{3.143} & \multicolumn{1}{|l}{3.092} & 
\multicolumn{1}{|l|}{2.982} \\ \cline{1-1}\cline{3-7}\cline{7-7}
\multicolumn{1}{|l}{BVS$_{{\rm La(Ba)}}$} & \multicolumn{1}{|l}{} & 
\multicolumn{1}{|l}{2.799} & \multicolumn{1}{|l}{2.848} & 
\multicolumn{1}{|l}{2.890} & \multicolumn{1}{|l}{2.844} & 
\multicolumn{1}{|l|}{2.797} \\ \cline{1-1}\cline{3-7}\cline{7-7}
\multicolumn{1}{|l}{BVS$_{{\rm Ba}}$} & \multicolumn{1}{|l}{} & 
\multicolumn{1}{|l}{2.024} & \multicolumn{1}{|l}{2.026} & 
\multicolumn{1}{|l}{2.040} & \multicolumn{1}{|l}{2.062} & 
\multicolumn{1}{|l|}{2.092} \\ \cline{1-1}\cline{3-7}
\end{tabular}

{\bf (b)}

\begin{tabular}{llllllllll}
\cline{1-1}\cline{3-10}\cline{6-6}
\multicolumn{1}{|l}{Ca conc. $x$} & \multicolumn{1}{|l}{} & 
\multicolumn{1}{|l}{} &  &  &  & 0.4 &  &  & \multicolumn{1}{l|}{} \\ 
\cline{1-1}\cline{3-10}\cline{6-6}\cline{10-10}
&  &  &  &  &  &  &  &  &  \\ \cline{1-1}\cline{3-10}\cline{10-10}
\multicolumn{1}{|l}{O conc. $y$} & \multicolumn{1}{|l}{} & 
\multicolumn{1}{|l}{6.884} & \multicolumn{1}{|l}{6.926} & 
\multicolumn{1}{|l}{7.008} & \multicolumn{1}{|l}{7.068} & 
\multicolumn{1}{|l}{7.142} & \multicolumn{1}{|l}{7.174} & 
\multicolumn{1}{|l}{7.208} & \multicolumn{1}{|l|}{7.240} \\ 
\cline{1-1}\cline{3-10}
\multicolumn{1}{|l}{BVS$_{{\rm Ca}}$} & \multicolumn{1}{|l}{} & 
\multicolumn{1}{|l}{1.865} & \multicolumn{1}{|l}{1.876} & 
\multicolumn{1}{|l}{1.882} & \multicolumn{1}{|l}{1.881} & 
\multicolumn{1}{|l}{1.882} & \multicolumn{1}{|l}{1.883} & 
\multicolumn{1}{|l}{1.876} & \multicolumn{1}{|l|}{1.854} \\ 
\cline{1-1}\cline{3-10}
\multicolumn{1}{|l}{BVS$_{{\rm La(Y)}}$} & \multicolumn{1}{|l}{} & 
\multicolumn{1}{|l}{3.245} & \multicolumn{1}{|l}{3.264} & 
\multicolumn{1}{|l}{3.275} & \multicolumn{1}{|l}{3.273} & 
\multicolumn{1}{|l}{3.276} & \multicolumn{1}{|l}{3.278} & 
\multicolumn{1}{|l}{3.265} & \multicolumn{1}{|l|}{3.226} \\ 
\cline{1-1}\cline{3-10}
\multicolumn{1}{|l}{BVS$_{{\rm Ba}}$} & \multicolumn{1}{|l}{} & 
\multicolumn{1}{|l}{2.121} & \multicolumn{1}{|l}{2.131} & 
\multicolumn{1}{|l}{2.142} & \multicolumn{1}{|l}{2.145} & 
\multicolumn{1}{|l}{2.143} & \multicolumn{1}{|l}{2.161} & 
\multicolumn{1}{|l}{2.159} & \multicolumn{1}{|l|}{2.147} \\ 
\cline{1-1}\cline{3-10}
\multicolumn{1}{|l}{BVS$_{{\rm La(Ba)}}$} & \multicolumn{1}{|l}{} & 
\multicolumn{1}{|l}{2.780} & \multicolumn{1}{|l}{2.821} & 
\multicolumn{1}{|l}{2.786} & \multicolumn{1}{|l}{2.797} & 
\multicolumn{1}{|l}{2.844} & \multicolumn{1}{|l}{2.898} & 
\multicolumn{1}{|l}{1.866} & \multicolumn{1}{|l|}{2.918} \\ 
\cline{1-1}\cline{3-3}\cline{3-10}\cline{5-5}\cline{7-7}\cline{9-10}
\end{tabular}

\bigskip \eject

\section{Table 8}

{\bf (a)}

\bigskip 
\begin{tabular}{ccccccc}
\cline{1-1}\cline{3-7}\cline{7-7}
\multicolumn{1}{|c}{Calcium conc. $x$} & \multicolumn{1}{|c}{} & 
\multicolumn{1}{|c}{} &  & 0.1 &  & \multicolumn{1}{c|}{} \\ 
\cline{1-1}\cline{3-7}
&  &  &  &  &  &  \\ \cline{1-1}\cline{3-7}
\multicolumn{1}{|c}{Oxygen conc. $y$} & \multicolumn{1}{|c}{} & 
\multicolumn{1}{|c}{7.038} & \multicolumn{1}{|c}{7.084} & 
\multicolumn{1}{|c}{7.158} & \multicolumn{1}{|c}{7.180} & 
\multicolumn{1}{|c|}{7.258} \\ \cline{1-1}\cline{3-7}
\multicolumn{1}{|c}{$T_{c}$ (K)} & \multicolumn{1}{|c}{} & 
\multicolumn{1}{|c}{30.9} & \multicolumn{1}{|c}{42.6} & \multicolumn{1}{|c}{
52.6} & \multicolumn{1}{|c}{41.4} & \multicolumn{1}{|c|}{5.0} \\ 
\cline{1-1}\cline{3-7}
\multicolumn{1}{|c}{$\xi $$_{{\rm CuI}}=V_{{\rm CuI}}^{{\rm Avg}}-2$} & 
\multicolumn{1}{|c}{} & \multicolumn{1}{|c}{0.094} & \multicolumn{1}{|c}{
0.154} & \multicolumn{1}{|c}{0.267} & \multicolumn{1}{|c}{0.312} & 
\multicolumn{1}{|c|}{0.417} \\ \cline{1-1}\cline{3-7}
\multicolumn{1}{|c}{$\xi $$_{{\rm CuII}}=V_{{\rm CuII}}^{{\rm Avg}}-2$} & 
\multicolumn{1}{|c}{} & \multicolumn{1}{|c}{0.114} & \multicolumn{1}{|c}{
0.121} & \multicolumn{1}{|c}{0.121} & \multicolumn{1}{|c}{0.140} & 
\multicolumn{1}{|c|}{0.151} \\ \cline{1-1}\cline{3-7}
\multicolumn{1}{|c}{$V_{{\rm Global}}^{{\rm Avg}}$} & \multicolumn{1}{|c}{}
& \multicolumn{1}{|c}{2.107} & \multicolumn{1}{|c}{2.132} & 
\multicolumn{1}{|c}{2.170} & \multicolumn{1}{|c}{2.198} & 
\multicolumn{1}{|c|}{2.240} \\ \cline{1-1}\cline{3-7}
\multicolumn{1}{|c}{$V_{{\rm Global}}^{{\rm Stoich.}}$} & 
\multicolumn{1}{|c}{} & \multicolumn{1}{|c}{2.275} & \multicolumn{1}{|c}{
2.306} & \multicolumn{1}{|c}{2.355} & \multicolumn{1}{|c}{2.370} & 
\multicolumn{1}{|c|}{2.422} \\ \cline{1-1}\cline{3-7}
\end{tabular}

{\bf (b)}

\begin{tabular}{cccccccccc}
\cline{1-1}\cline{3-10}\cline{7-10}
\multicolumn{1}{|c}{Calcium conc. $x$} & \multicolumn{1}{|c}{} & 
\multicolumn{1}{|c}{} &  &  & 0.4 &  &  &  & \multicolumn{1}{c|}{} \\ 
\cline{1-1}\cline{3-10}
&  &  &  &  &  &  &  &  &  \\ \cline{1-1}\cline{3-10}
\multicolumn{1}{|c}{Oxygen conc. $y$} & \multicolumn{1}{|c}{} & 
\multicolumn{1}{|c}{6.884} & \multicolumn{1}{|c}{6.926} & 
\multicolumn{1}{|c}{7.008} & \multicolumn{1}{|c}{7.068} & 
\multicolumn{1}{|c}{7.142} & \multicolumn{1}{|c}{7.174} & 
\multicolumn{1}{|c}{7.208} & \multicolumn{1}{|c|}{7.240} \\ 
\cline{1-1}\cline{3-10}\cline{7-10}
\multicolumn{1}{|c}{$T_{c}$ (K)} & \multicolumn{1}{|c}{} & 
\multicolumn{1}{|c}{12.6} & \multicolumn{1}{|c}{33.1} & \multicolumn{1}{|c}{
54.6} & \multicolumn{1}{|c}{71.4} & \multicolumn{1}{|c}{79.4} & 
\multicolumn{1}{|c}{80.3} & \multicolumn{1}{|c}{75.9} & \multicolumn{1}{|c|}{
60.8} \\ \cline{1-1}\cline{3-10}\cline{7-10}
\multicolumn{1}{|c}{$\xi $$_{{\rm CuI}}=V_{{\rm CuI}}^{{\rm Avg}}-2$} & 
\multicolumn{1}{|c}{} & \multicolumn{1}{|c}{0.048} & \multicolumn{1}{|c}{
0.079} & \multicolumn{1}{|c}{0.176} & \multicolumn{1}{|c}{0.264} & 
\multicolumn{1}{|c}{0.346} & \multicolumn{1}{|c}{0.385} & 
\multicolumn{1}{|c}{0.420} & \multicolumn{1}{|c|}{0.433} \\ 
\cline{1-1}\cline{3-10}\cline{7-10}
\multicolumn{1}{|c}{$\xi $$_{{\rm CuII}}=V_{{\rm CuII}}^{{\rm Avg}}-2$} & 
\multicolumn{1}{|c}{} & \multicolumn{1}{|c}{0.167} & \multicolumn{1}{|c}{
0.180} & \multicolumn{1}{|c}{0.198} & \multicolumn{1}{|c}{0.214} & 
\multicolumn{1}{|c}{0.222} & \multicolumn{1}{|c}{0.237} & 
\multicolumn{1}{|c}{0.246} & \multicolumn{1}{|c|}{0.246} \\ 
\cline{1-1}\cline{3-10}
\multicolumn{1}{|c}{$V_{{\rm Global}}^{{\rm Avg}}$} & \multicolumn{1}{|c}{}
& \multicolumn{1}{|c}{2.127} & \multicolumn{1}{|c}{2.146} & 
\multicolumn{1}{|c}{2.191} & \multicolumn{1}{|c}{2.231} & 
\multicolumn{1}{|c}{2.263} & \multicolumn{1}{|c}{2.286} & 
\multicolumn{1}{|c}{2.304} & \multicolumn{1}{|c|}{2.308} \\ 
\cline{1-1}\cline{3-10}
\multicolumn{1}{|c}{$V_{{\rm Global}}^{{\rm Stoich.}}$} & 
\multicolumn{1}{|c}{} & \multicolumn{1}{|c}{2.173} & \multicolumn{1}{|c}{
2.201} & \multicolumn{1}{|c}{2.255} & \multicolumn{1}{|c}{2.295} & 
\multicolumn{1}{|c}{2.345} & \multicolumn{1}{|c}{2.366} & 
\multicolumn{1}{|c}{2.389} & \multicolumn{1}{|c|}{2.410} \\ 
\cline{1-1}\cline{3-10}
\end{tabular}
\strut \mathstrut

\medskip \eject

\section{Table 9}

(a)

\begin{tabular}{lllllll}
\cline{1-1}\cline{3-7}\cline{7-7}
\multicolumn{1}{|l}{Ca conc. $x$} & \multicolumn{1}{|l}{} & 
\multicolumn{1}{|l}{} &  & 0.1 &  & \multicolumn{1}{l|}{} \\ 
\cline{1-1}\cline{3-7}\cline{7-7}
&  &  &  &  &  &  \\ \cline{1-1}\cline{3-7}
\multicolumn{1}{|l}{$T_{c}$} & \multicolumn{1}{|l}{} & \multicolumn{1}{|l}{
30.9} & \multicolumn{1}{|l}{42.6} & \multicolumn{1}{|l}{52.6} & 
\multicolumn{1}{|l}{41.4} & \multicolumn{1}{|l|}{5.0} \\ 
\cline{1-1}\cline{3-7}
\multicolumn{1}{|l}{O conc. $y$} & \multicolumn{1}{|l}{} & 
\multicolumn{1}{|l}{7.038} & \multicolumn{1}{|l}{7.084} & 
\multicolumn{1}{|l}{7.158} & \multicolumn{1}{|l}{7.180} & 
\multicolumn{1}{|l|}{7.258} \\ 
\cline{1-1}\cline{3-3}\cline{3-7}\cline{5-5}\cline{7-7}
\multicolumn{1}{|l}{BVS$_{{\rm Op}}$} & \multicolumn{1}{|l}{} & 
\multicolumn{1}{|l}{2.073} & \multicolumn{1}{|l}{2.068} & 
\multicolumn{1}{|l}{2.056} & \multicolumn{1}{|l}{2.043} & 
\multicolumn{1}{|l|}{2.017} \\ 
\cline{1-1}\cline{3-3}\cline{3-7}\cline{5-5}\cline{7-7}
\multicolumn{1}{|l}{BVS$_{{\rm O\alpha (1)}}$} & \multicolumn{1}{|l}{} & 
\multicolumn{1}{|l}{1.659} & \multicolumn{1}{|l}{1.694} & 
\multicolumn{1}{|l}{1.725} & \multicolumn{1}{|l}{1.774} & 
\multicolumn{1}{|l|}{1.835} \\ \cline{1-1}\cline{3-7}
\multicolumn{1}{|l}{BVS$_{{\rm O\alpha (2)}}$} & \multicolumn{1}{|l}{} & 
\multicolumn{1}{|l}{1.823} & \multicolumn{1}{|l}{1.831} & 
\multicolumn{1}{|l}{1.894} & \multicolumn{1}{|l}{1.874} & 
\multicolumn{1}{|l|}{1.765} \\ \cline{1-1}\cline{3-7}\cline{7-7}
\multicolumn{1}{|l}{BVS$_{{\rm Oc(1)}}$} & \multicolumn{1}{|l}{} & 
\multicolumn{1}{|l}{1.886} & \multicolumn{1}{|l}{1.905} & 
\multicolumn{1}{|l}{1.923} & \multicolumn{1}{|l}{1.942} & 
\multicolumn{1}{|l|}{1.962} \\ \cline{1-1}\cline{3-3}\cline{3-7}\cline{4-7}
\multicolumn{1}{|l}{BVS$_{{\rm Oc(2)}}$} & \multicolumn{1}{|l}{} & 
\multicolumn{1}{|l}{1.889} & \multicolumn{1}{|l}{1.907} & 
\multicolumn{1}{|l}{1.916} & \multicolumn{1}{|l}{1.924} & 
\multicolumn{1}{|l|}{1.921} \\ \cline{1-1}\cline{3-3}\cline{3-7}\cline{4-6}
\multicolumn{1}{|l}{BVS$_{{\rm Oc(3)}}$} & \multicolumn{1}{|l}{} & 
\multicolumn{1}{|l}{1.842} & \multicolumn{1}{|l}{1.852} & 
\multicolumn{1}{|l}{1.893} & \multicolumn{1}{|l}{1.902} & 
\multicolumn{1}{|l|}{1.920} \\ \cline{1-1}\cline{3-7}
\multicolumn{1}{|l}{$p_{{\rm Brown}}$} & \multicolumn{1}{|l}{} & 
\multicolumn{1}{|l}{0.284} & \multicolumn{1}{|l}{0.294} & 
\multicolumn{1}{|l}{0.298} & \multicolumn{1}{|l}{0.319} & 
\multicolumn{1}{|l|}{0.334} \\ \cline{1-1}\cline{3-7}
\multicolumn{1}{|l}{$p_{{\rm Tallon}}^{\xi \neq 0}$} & \multicolumn{1}{|l}{}
& \multicolumn{1}{|l}{-0.031} & \multicolumn{1}{|l}{-0.015} & 
\multicolumn{1}{|l}{0.008} & \multicolumn{1}{|l}{0.055} & 
\multicolumn{1}{|l|}{0.117} \\ \cline{1-1}\cline{3-7}
\end{tabular}

(b)

\begin{tabular}{llllllllll}
\cline{1-1}\cline{3-10}\cline{6-6}
\multicolumn{1}{|l}{Ca conc. $x$} & \multicolumn{1}{|l}{} & 
\multicolumn{1}{|l}{} &  &  &  & 0.4 &  &  & \multicolumn{1}{l|}{} \\ 
\cline{1-1}\cline{3-10}\cline{6-6}\cline{10-10}
&  &  &  &  &  &  &  &  &  \\ \cline{1-1}\cline{3-10}\cline{4-10}
\multicolumn{1}{|l}{$T_{c}$} & \multicolumn{1}{|l}{} & \multicolumn{1}{|l}{
12.6} & \multicolumn{1}{|l}{33.1} & \multicolumn{1}{|l}{54.6} & 
\multicolumn{1}{|l}{71.4} & \multicolumn{1}{|l}{79.4} & \multicolumn{1}{|l}{
80.3} & \multicolumn{1}{|l}{75.9} & \multicolumn{1}{|l|}{60.8} \\ 
\cline{1-1}\cline{3-10}\cline{4-10}
\multicolumn{1}{|l}{O conc. $y$} & \multicolumn{1}{|l}{} & 
\multicolumn{1}{|l}{6.884} & \multicolumn{1}{|l}{6.926} & 
\multicolumn{1}{|l}{7.008} & \multicolumn{1}{|l}{7.068} & 
\multicolumn{1}{|l}{7.148} & \multicolumn{1}{|l}{7.174} & 
\multicolumn{1}{|l}{7.208} & \multicolumn{1}{|l|}{7.240} \\ 
\cline{1-1}\cline{3-10}
\multicolumn{1}{|l}{BVS$_{{\rm Op}}$} & \multicolumn{1}{|l}{} & 
\multicolumn{1}{|l}{2.151} & \multicolumn{1}{|l}{2.154} & 
\multicolumn{1}{|l}{2.142} & \multicolumn{1}{|l}{2.131} & 
\multicolumn{1}{|l}{2.121} & \multicolumn{1}{|l}{2.127} & 
\multicolumn{1}{|l}{2.113} & \multicolumn{1}{|l|}{2.099} \\ 
\cline{1-1}\cline{3-10}
\multicolumn{1}{|l}{BVS$_{{\rm O\alpha }\left( 1\right) }$} & 
\multicolumn{1}{|l}{} & \multicolumn{1}{|l}{1.538} & \multicolumn{1}{|l}{
1.555} & \multicolumn{1}{|l}{1.597} & \multicolumn{1}{|l}{1.841} & 
\multicolumn{1}{|l}{1.673} & \multicolumn{1}{|l}{1.691} & 
\multicolumn{1}{|l}{1.718} & \multicolumn{1}{|l|}{1.731} \\ 
\cline{1-1}\cline{3-10}
\multicolumn{1}{|l}{BVS$_{{\rm O\alpha }\left( 2\right) }$} & 
\multicolumn{1}{|l}{} & \multicolumn{1}{|l}{1.852} & \multicolumn{1}{|l}{
1.890} & \multicolumn{1}{|l}{1.888} & \multicolumn{1}{|l}{1.920} & 
\multicolumn{1}{|l}{1.931} & \multicolumn{1}{|l}{1.948} & 
\multicolumn{1}{|l}{1.944} & \multicolumn{1}{|l|}{1.925} \\ 
\cline{1-1}\cline{3-10}
\multicolumn{1}{|l}{BVS$_{\text{{\rm O}{\it c(1}}{\rm )}}$} & 
\multicolumn{1}{|l}{} & \multicolumn{1}{|l}{1.864} & \multicolumn{1}{|l}{
1.874} & \multicolumn{1}{|l}{1.893} & \multicolumn{1}{|l}{1.918} & 
\multicolumn{1}{|l}{1.933} & \multicolumn{1}{|l}{1.941} & 
\multicolumn{1}{|l}{1.952} & \multicolumn{1}{|l|}{1.950} \\ 
\cline{1-1}\cline{3-3}\cline{3-10}\cline{5-5}\cline{7-7}\cline{9-10}
\multicolumn{1}{|l}{BVS$_{\text{{\rm Oc(2)}}}$} & \multicolumn{1}{|l}{} & 
\multicolumn{1}{|l}{1.924} & \multicolumn{1}{|l}{1.929} & 
\multicolumn{1}{|l}{1.934} & \multicolumn{1}{|l}{1.951} & 
\multicolumn{1}{|l}{1.958} & \multicolumn{1}{|l}{1.961} & 
\multicolumn{1}{|l}{1.965} & \multicolumn{1}{|l|}{1.970} \\ 
\cline{1-1}\cline{3-3}\cline{3-10}\cline{5-5}\cline{7-7}\cline{9-10}
\multicolumn{1}{|l}{BVS$_{\text{{\rm O}{\it c}{\rm (3)}}}$} & 
\multicolumn{1}{|l}{} & \multicolumn{1}{|l}{1.884} & \multicolumn{1}{|l}{
1.906} & \multicolumn{1}{|l}{1.904} & \multicolumn{1}{|l}{1.913} & 
\multicolumn{1}{|l}{1,922} & \multicolumn{1}{|l}{1.959} & 
\multicolumn{1}{|l}{1.944} & \multicolumn{1}{|l|}{1.934} \\ 
\cline{1-1}\cline{3-10}
\multicolumn{1}{|l}{$p_{{\rm Brown}}$} & \multicolumn{1}{|l}{} & 
\multicolumn{1}{|l}{0.213} & \multicolumn{1}{|l}{0.231} & 
\multicolumn{1}{|l}{0.261} & \multicolumn{1}{|l}{0.284} & 
\multicolumn{1}{|l}{0.302} & \multicolumn{1}{|l}{0.322} & 
\multicolumn{1}{|l}{0.335} & \multicolumn{1}{|l|}{0.339} \\ 
\cline{1-1}\cline{3-10}
\multicolumn{1}{|l}{$p_{{\rm Tallon}}^{\xi \neq 0}$} & \multicolumn{1}{|l}{}
& \multicolumn{1}{|l}{-0.135} & \multicolumn{1}{|l}{-0.127} & 
\multicolumn{1}{|l}{-0.086} & \multicolumn{1}{|l}{-0.049} & 
\multicolumn{1}{|l}{-0.020} & \multicolumn{1}{|l}{-0.017} & 
\multicolumn{1}{|l}{0.019} & \multicolumn{1}{|l|}{0.048} \\ 
\cline{1-1}\cline{3-10}
\end{tabular}

\bigskip

\section{Table 10}

\begin{tabular}{|c|c|c|c|c|c|}
\cline{1-1}\cline{3-6}
Oxygen conc. $y$ &  & 6.95 & 6.84 & 6.81 & 6.78 \\ \cline{1-1}\cline{3-6}
$T_{c}$ &  & 90 & 88 & 86 & 80 \\ \cline{1-1}\cline{3-6}
$p_{{\rm Brown}}$ &  & 0.264 & 0.238 & 0.222 & 0.207 \\ 
\cline{1-1}\cline{3-6}
$p_{{\rm Tallon}}^{\xi \neq 0}$ &  & 0.105 & 0.089 & 0.075 & 0.068 \\ 
\cline{1-1}\cline{3-6}
$\ p_{{\rm Tallon}}^{\xi =0}$ &  & 0.165 & 0.150 & 0.135 & 0.125 \\ 
\cline{1-1}\cline{3-6}
${\cal C}_{{\rm Brown}}$(\%) &  & 53.9 & 58.8 & 69.8 & 71.3 \\ 
\cline{1-1}\cline{3-6}
${\cal C}_{{\rm Tallon}}^{\xi \neq 0}$ (\%) &  & 23.0 & 23.3 & 26.1 & 24.3
\\ \cline{1-1}\cline{3-6}
${\cal C}_{{\rm Tallon}}^{\xi =0}$(\%) &  & 36.7 & 44.1 & 43.4 & 44.6 \\ 
\cline{1-1}\cline{3-6}
\end{tabular}

\section{ Acknowledgements}

\bigskip We are indebted to Professor I.D. Brown for valuable criticism.

This research was supported by the Israel Science Foundation (administered
by the Israel Academy of Sciences and Humanities), by the Center of
Absorption in Science (Ministry of Immigrant Absorption, State of Israel),
and by the fund for Promotion of Research at the Technion. \ \ One of us
(O.C.) \ acknowledges support from ARPA/ONR and the State of Illinois under
HECA.

\section{\bf Table captions}

Table 1: Values of the BVS parameter $r_{0}$\ for all the ions relevant to
the present study\cite{brown-alt}.

Table2$:$ \ This is a specimen table, giving the coordinates and atomic
displacement parameters of the ions in the unit cell, from the na\"{i}ve
Rietveld refinement when all ions are held to their ``ideal'' positions\cite
{table}. \ The full table is available by e-mail from
charles@physics.technion.ac.il

Table 3: The na\"\i ve interionic distances, calculated by Rietveld
refinement, when the O$_{c}$and O$_{\alpha }$ are not allowed to deviate
from their ``ideal'' positions.

Table 4: \ Na\"\i ve bond valence sums for Ca, La, Ba, and O. \ Note the
extremely poor values for La$_{{\rm (Ba)}}$, indicating the presence of
significant distortion away from the ``ideal'' structure. \ The oxygen BVS's
require the use of the mixing ratios $\xi $, which are introduced in \S 5.

Table 5: \ This specimen table\cite{table} replaces Table 2, giving the
coordinates and atomic displacement parameters of the ions in the unit cell,
from the Rietveld refinement when the O$_{c}$ \ and O$_{\alpha }$ \ ions are
allowed to deviate from their ``ideal'' positions\cite{table}. \ The full
table is available by e-mail from charles@physics.technion.ac.il

Table 6: \ The distances in this table are calculated from the Rietveld
refinement, when the O$_{c}$and O$_{\alpha }$ are allowed to deviate from
their ``ideal'' positions. \ The suffix 1 refers to the undisplaced ion, and
the suffixes 2 and 3 to the displaced one. \ Note that for the situations
described in Fig. 2 D and E, \ there are two distinct distances of Ba and of
La$_{{\rm (Ba)}}$ to O$_{c(2)},$ corresponding to displacement {\it along}
the line joining La to Ba (Case $r_{\parallel }$) and to displacement {\it %
orthogonal }to the line joining the pair of similar ions (Case $r_{\perp }$).

Table 7: \ The BVS's of Ca, La, and Ba, recalculated using the ``improved''
Rietveld refinements from Table 6.

Table 8: \ Mixing ratios \ $\xi $$_{{\rm I}}$ and $\xi $$_{{\rm II}}$; and
comparison of the global average Cu valence from BVS and from stoichiometry.

Table 9: \ The BVS's of all the oxygen ions, and the values of $p_{{\rm Brown%
}}$ and $p_{{\rm Tallon}}^{\xi \neq 0}$ for CLBLCO. \ Note that although 
{\it most }of the oxygen BVS's are quite close to 2, those of O$_{\alpha }$
are rather poor, reflecting (a) the likelihood that some lattice distortion
arises, due to the fact that there are many oxygen vacancies in the CuI
layer, and (b) the possibility that at least some of the O$_{\alpha }$ ions
do not lie exactly in the plane ({\it i.e.} that the dimerization of the
La.\ ions is incomplete).

Table 10:\ The values, for YBCO, of $p_{{\rm Brown}}$, $p_{{\rm Tallon}%
}^{\xi \neq 0}$, and $\ p_{{\rm Tallon}}^{\xi =0}$ are taken from ref. [16].
\ We have not included values of \ $y$\ less than 6.64, since they enter the
region where $\xi ^{({\rm I})}$ is negative, {\it i.e. }the CuI layer
contains electrons rather than holes. \ The charge concentration on the CuII
layers is ${\cal C=}2p/(V_{{\rm Global}}^{{\rm Stoich.}}-2)$.

\section{\bf Figure Captions}

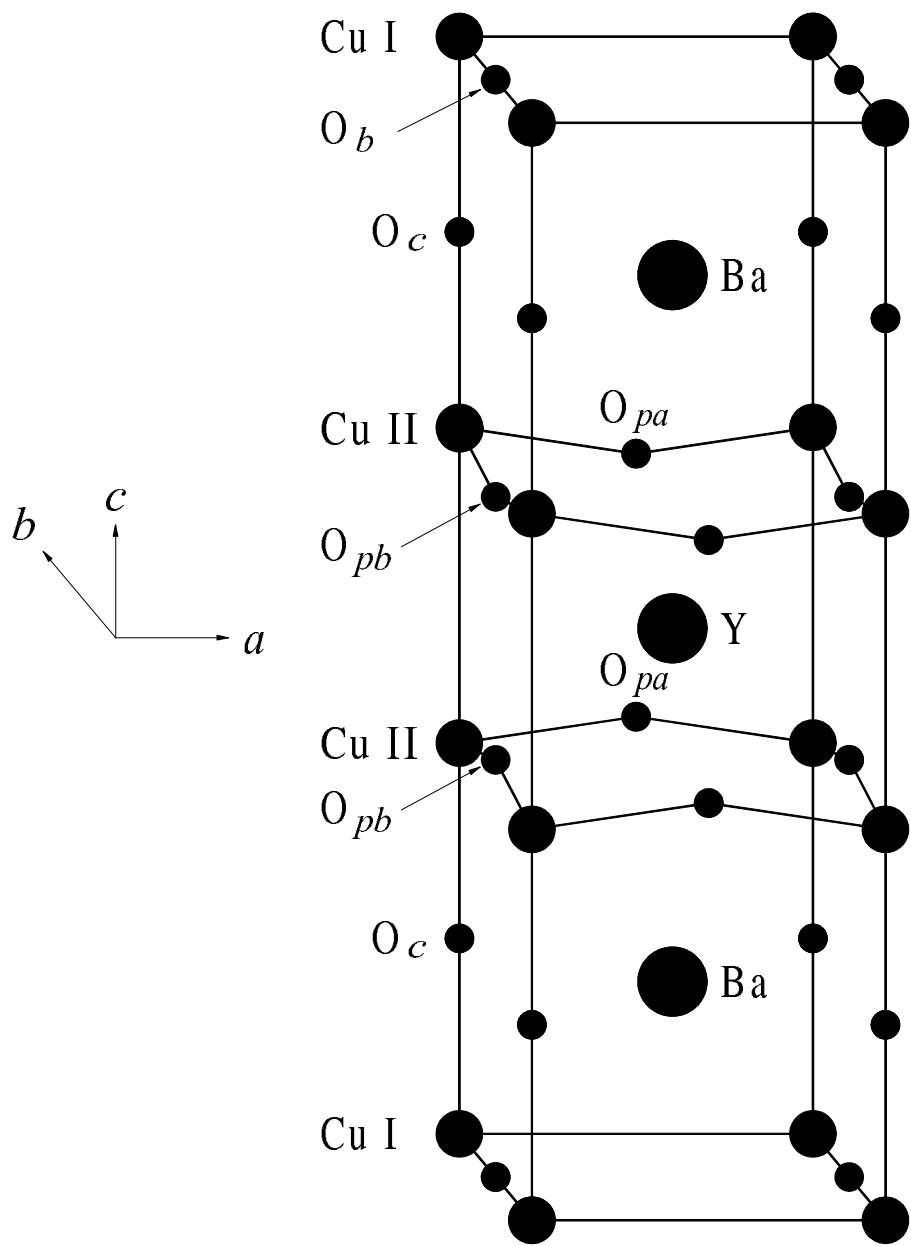
Fig. 1: Unit cell of YBCO, defining our labellng convention.. \ CLBLCO is
always tetragonal; since there are no chains, the $a$\ and\ $b$\ directions
are equivalent. \ In \ the CuII layers, O$_{pa}$\ \ and O$_{pb}$\ \ need not
be distinguished, and will be labeled O$_{p}$.\ The oxygens in the CuI layer
occupy the O$_{b}$\ and O$_{a}$\ \ sites with equal probability and are
assumed to be distributed randomly; we will label them O$_{{\rm \alpha }}$.
\ 

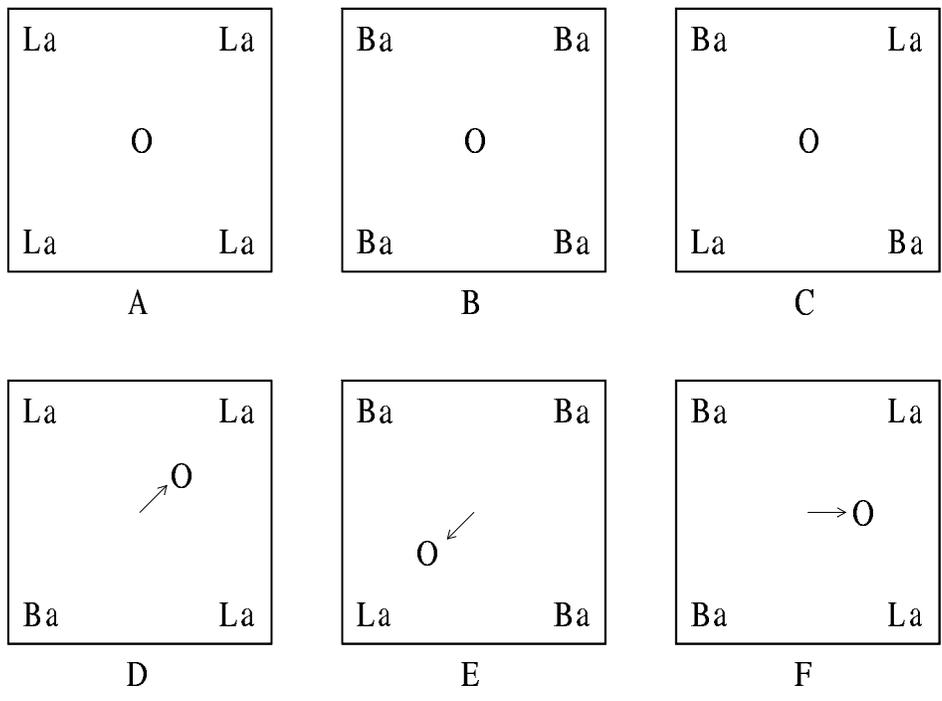
Fig. 2:\ The six distinct environments of an O$_{c}$ atom, showing the
directions of possible displacement. \ In cases A, B, and C, by symmetry,
the oxygen is not displaced from its ``ideal'' position. In case D, it is
displaced towards the solitary La, in case E, away from the solitary Ba, and
in case F, towards the La pair. Cases A and B are nondegenerate, C is doubly
degenerate, and D, E, and F are fourfold degenerate.

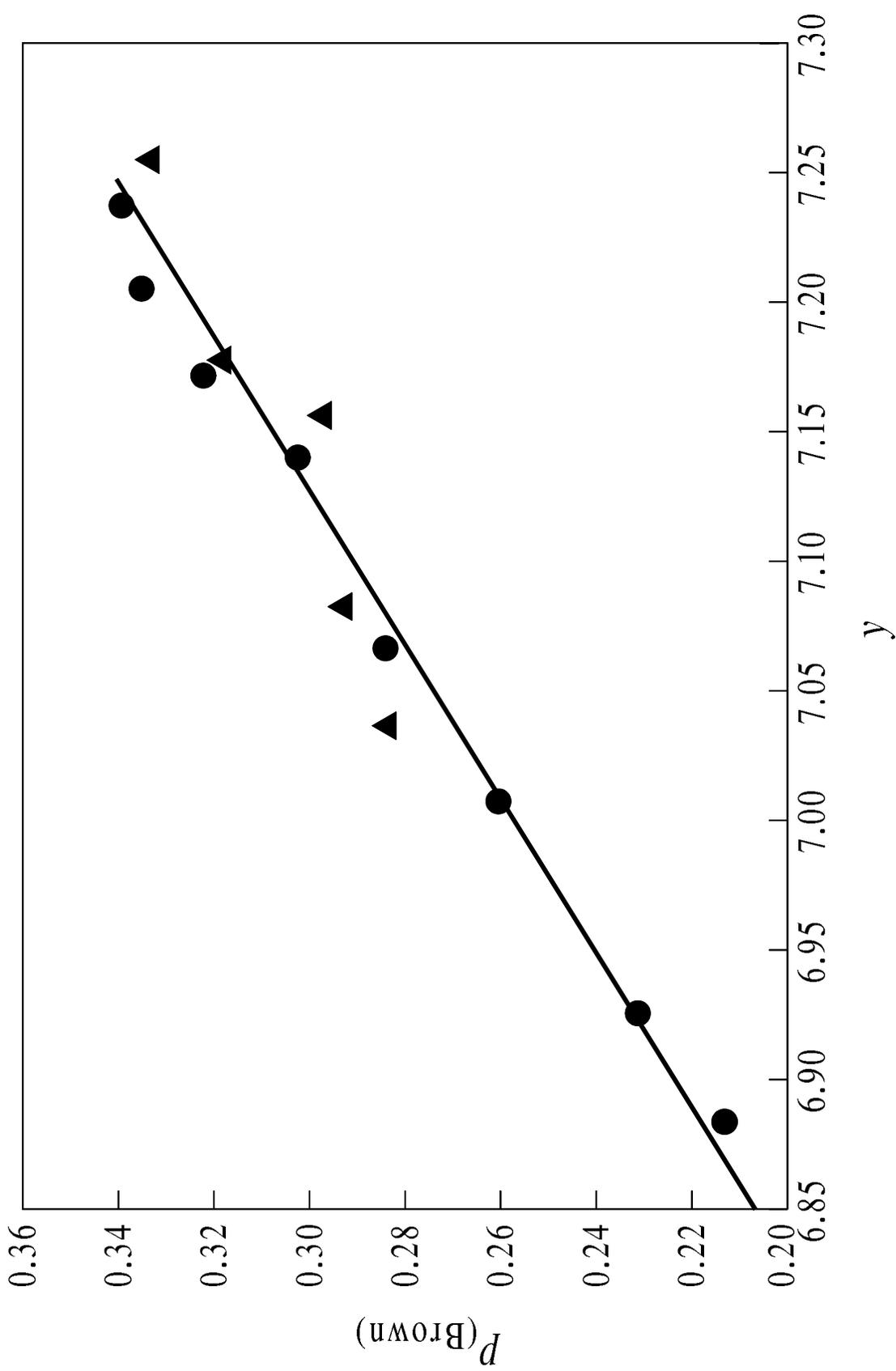
Fig. 3: CLBLCO --- Plot of $p_{{\rm Brown}}$ versus oxygen concentration $y$%
, for nominal calcium concentration $x=0.4$\ (squares) and $x=0.1$\ (circles)%
$.$

Fig. 4: CLBLCO --- (a) Plot of $T_{c}$ versus $p_{{\rm Brown}}$, and (b)
plot of $T_{c}$\ \ versus $y,$ for nominal $x=0.4$ (squares) and $x=0.1$\
(circles)$.$\

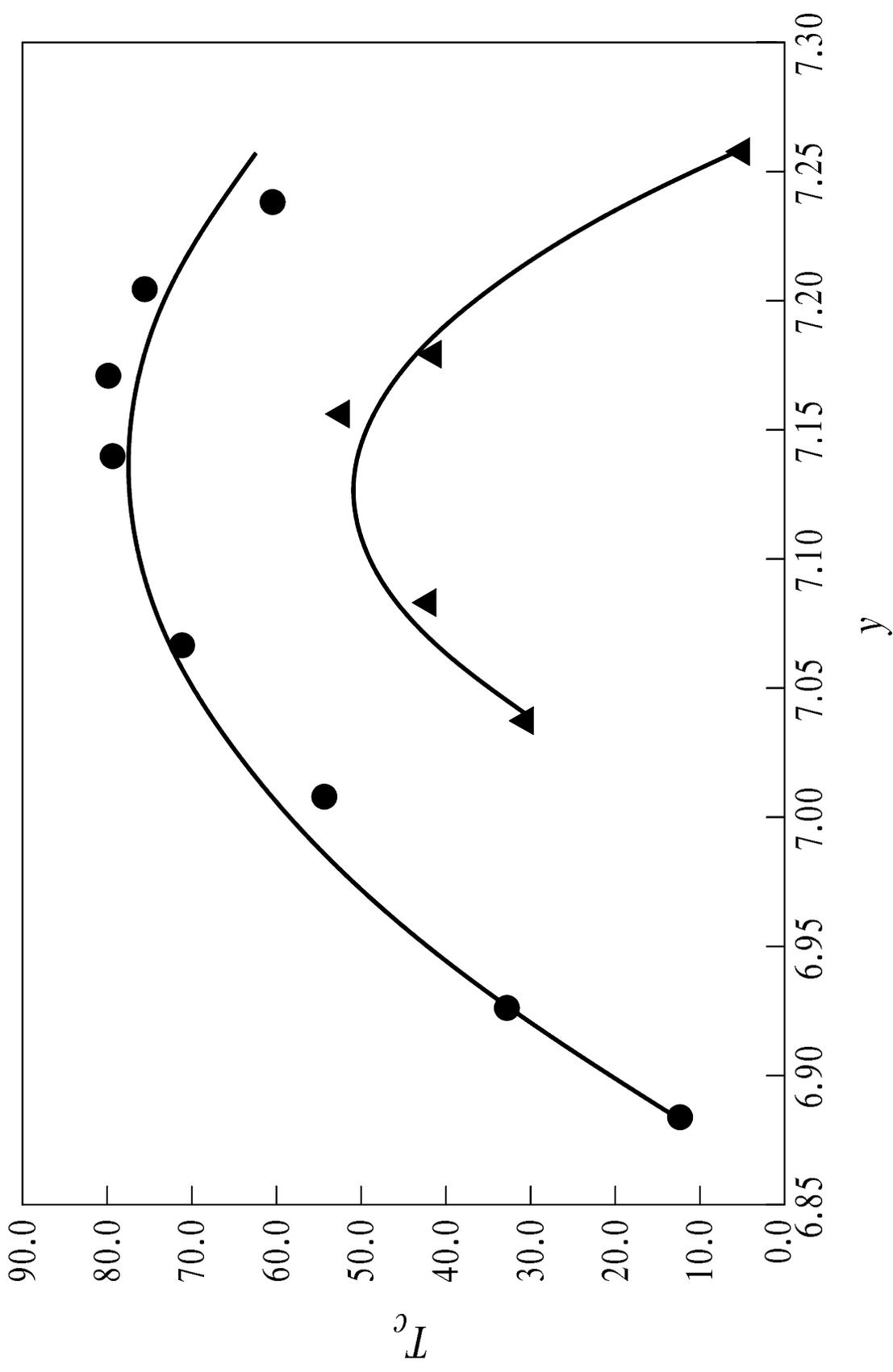
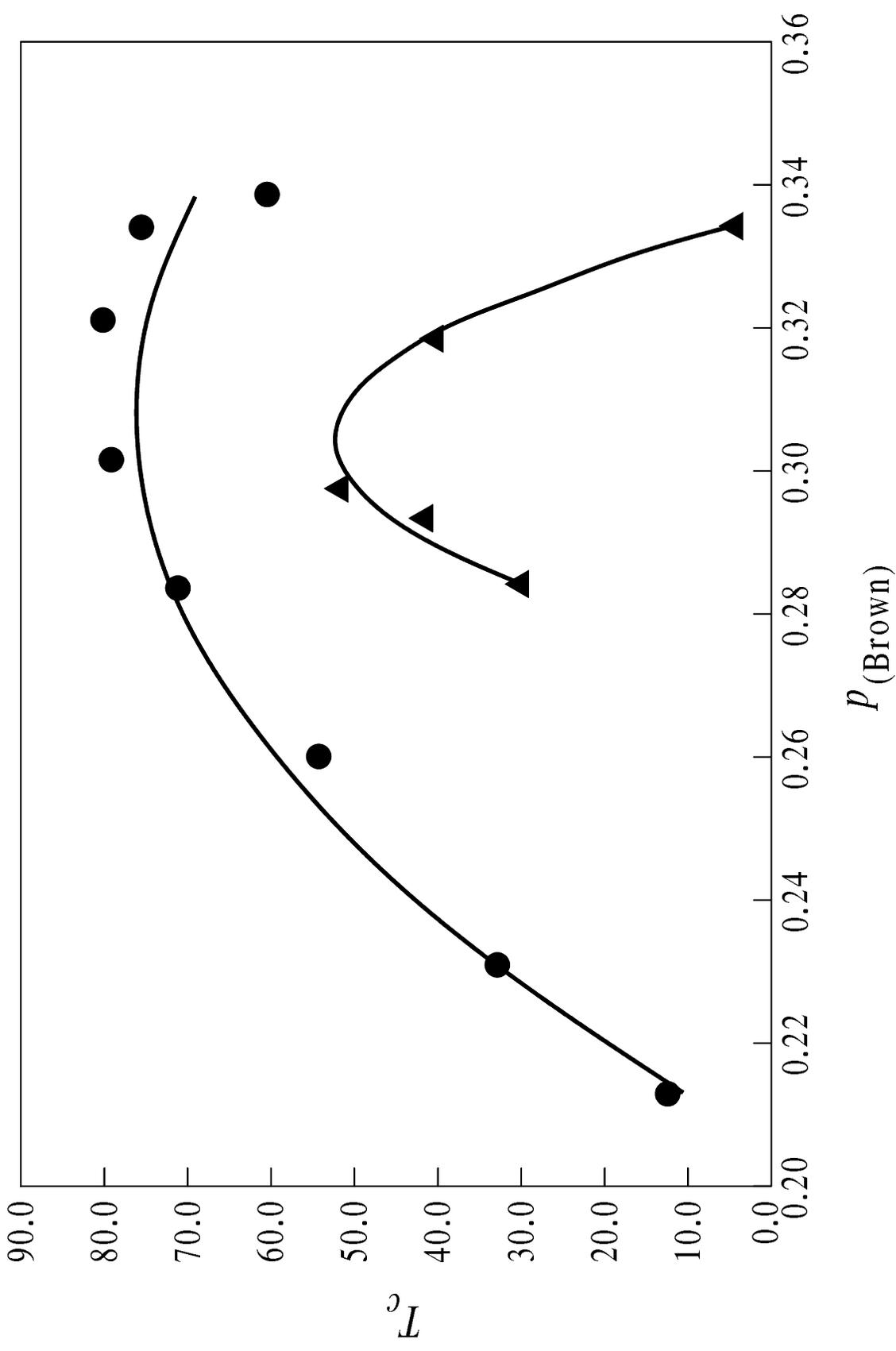
\end{document}